\documentclass[]{aa}  
\usepackage{graphicx}
\usepackage{txfonts}
\usepackage{graphicx}
\usepackage{tabularx}
\usepackage{amsmath}
\usepackage{rotating}
\usepackage{xcolor}
\usepackage[]{hyperref}
\usepackage{float}
\usepackage{morefloats}
\usepackage{natbib}
\usepackage{url}
\usepackage{times}
\def\spose#1{\hbox to 0pt{#1\hss}}
\newcommand\lsim{\mathrel{\spose{\lower 3pt\hbox{$\mathchar"218$}}
     \raise 2.0pt\hbox{$\mathchar"13C$}}}
\newcommand\gsim{\mathrel{\spose{\lower 3pt\hbox{$\mathchar"218$}}
     \raise 2.0pt\hbox{$\mathchar"13E$}}}

\makeatletter
\renewcommand*\aa@pageof{, page \thepage{} of \pageref*{LastPage}}
\makeatother

\begin{document} 

\title{Estimating black hole masses: Accretion disk fitting versus reverberation mapping and single epoch}
	\titlerunning{AD Fitting vs Reverberation Mapping \& Single Epoch}
\author{Samuele Campitiello\inst{1,2}\thanks{\email{sam.campitiello@gmail.com}} \and Annalisa Celotti\inst{1,3,4}
\and Gabriele Ghisellini\inst{3}
\and Tullia Sbarrato\inst{5}}

\institute{SISSA, Via Bonomea 265, I-34136, Trieste, Italy
   \and INAF - Osservatorio Astronomico di Trieste, Via Tiepolo 11, I-34131, Trieste, Italy \and INAF - Osservatorio Astronomico di Brera, via E. Bianchi 46, I-23807, Merate, Italy
\and INFN–Sezione di Trieste, via Valerio 2, I-34127 Trieste, Italy \and
Dipartimento di Fisica "G. Occhialini", Università di Milano - Bicocca, Piazza della Scienza 3, I-20126 Milano, Italy
            }

\date{Received ; accepted }

\abstract{We selected a sample of 28 Type 1 Active Galactic Nuclei for which a black hole mass has been inferred using the reverberation mapping technique and single epoch scaling relations. All 28 sources show clear evidence of the "Big Blue Bump" in the optical-UV band whose emission is produced by an accretion disk (AD) around a supermassive black hole. We fitted the spectrum of these sources with the relativistic thin AD model KERRBB in order to infer the black hole masses and compared them with those from Reverberation mapping and Single epoch methods, discussing the possible uncertainties linked to such a model by quantifying their weight on our results. We find that for the majority of the sources, KERRBB is a good description of the AD emission for a wide wavelength range. The overall uncertainty on the black hole mass estimated through the disk fitting procedure is $\sim 0.45$ dex (which includes the uncertainty on fitting parameters such as e.g. spin and viewing angle), comparable to the systematic uncertainty of reverberation mapping and single epoch methods; however, such an uncertainty can be $\lsim 0.3$ dex if one of the parameters of the fit is well constrained. Although all of the estimates are affected by large uncertainties, the masses inferred using the three methods are compatible if the dimensionless scale factor $f$ (linked to the unknown kinematics and geometry of the Broad Line Region) is assumed to be larger than one. For the majority of the sources, the comparison between the results coming from the three methods favors small spin values. To check the goodness of the KERRBB results, we compared them with those inferred with other models, such as AGNSED, a model that also accounts for the emission originating from an X-ray corona: using two sources with a good data coverage in the X band, we find that the masses estimated with the two models differ at most by a factor of $\sim 0.2$ dex.
}

\maketitle

%-------------------------------------------------------------------

\section{Introduction} \label{sec:intro}

Supermassive black holes (SMBHs) are located at the center of massive galaxies and determining their mass and spin is crucial for understanding their physical nature, the link with the host galaxy and their possible evolution in time.

Different methods have been used to estimate the mass of black holes (BHs) in Active Galactic Nuclei (AGNs): reverberation mapping ({\rm RM}, e.g., \citealt{Blandfo}; \citealt{Peterson}; \citealt{NetPet}; \citealt{Wanetal}; \citealt{Kaspietal}; \citealt{PetersonFerr}; \citealt{Benetal}; \citealt{Faus}; \citealt{BenMan}; \citealt{Shenetal19}; 2D velocity-delay maps, \citealt{Grier2}), single epoch ({\rm SE}) virial mass (e.g., \citealt{Vester}; \citealt{MclureJar}; \citealt{MclureDun}; \citealt{GreenHo}; \citealt{VesterPeter}; \citealt{OnKoll}; \citealt{Wangetal}; \citealt{VesterOs}; \citealt{Greene}; \citealt{Shenetal11}; \citealt{ShenLiu}; \citealt{TraNet}), AD fitting (e.g., \citealt{Malkan}; \citealt{SunMalk}; \citealt{WanPetr}; \citealt{Laor}; \citealt{Rok}; \citealt{Tripp}; \citealt{Ghi2010}; \citealt{Caldero}; \citealt{Campiti, Campiti_ar}, hereafter C18 and C19, respectively), microlensing in gravitationally lensed quasars (QSOs, e.g., \citealt{Irwin}; \citealt{Lewis}; \citealt{Rich04}; \citealt{Dai}; \citealt{MosKoc}; \citealt{Sluse}; \citealt{Guerras}), polarization in broad emission lines (e.g., \citealt{Savic}), dynamical BH mass (e.g., \citealt{Davies}; \citealt{Onkenetal07}; \citealt{HicMal}), and the recent method based on the redshift of Fe III lines in QSOs (\citealt{Media18, Media}). 

Each of them carries some uncertainties linked to the features of the system, to the parameters of the model involved for the estimates (e.g. \citealt{Laor}; \citealt{MclureJar}; \citealt{VesterPeter}; \citealt{Marconietal}; \citealt{Peterson10}; \citealt{Caldero}) and clearly to the quality of data. Therefore, in order to assess the robustness of the mass estimate, it is necessary to compare the results of the different methods and possibly to calibrate the model-scaling parameters. However, this is not trivial because different approaches are based on different observables and are thus not applicable to all sources.

The aim of this work is indeed to compare the results of the AD fitting method with those of RM in order to test the reliability of AD models as an alternative approach to evaluating SMBH masses in AGNs. We chose to compare our AD fitting results with those from RM because this latter is the most accurate technique based on direct measurements related to the Broad Line Region (BLR). As a supplementary test, we also include the SE results although the method is calibrated on RM measurements. 
 
The interpretation of the so-called "Big Blue Bump" component in the rest frame 1000 - 5000\ \AA\ as a "standard" AD emission has been questioned on various grounds (e.g. \citealt{Kora}; \citealt{Net2015}). One of the debated issues, particularly relevant for this work, is related to the presence of a soft X-ray excess component observed in many objects. Some authors have developed models to account for this excess in a self-consistent way (e.g., relativistically blurred photoionized disc reflection model, \citealt{RosFab}; \citealt{Crummy}; comptonization by an X-ray corona, e.g., \citealt{Done}; \citealt{KuDo} - hereafter KD18 - and references therein). Although the origin of the soft X-ray excess is not well established (e.g. \citealt{Caba}), in this paper we also discuss the possible effect of a warm corona above the disk that could modify the observed disk emission and therefore the determination of the BH mass. 

Despite numerous criticisms, the AD fitting procedure has been widely used in the literature and shows a general agreement with the Spectral Energy Distribution (SED) and in some cases with SE BH mass estimates (e.g., \citealt{DavLao}; \citealt{LaoDav}; \citealt{Caldero}; \citealt{Castietal}; \citealt{Capel1, Capel3}; \citealt{Maj16}; KD18; \citealt{Marcule}).

Although the aforementioned soft X-ray excess issue is still under debate, for a correct description of the AD emission, the more compelling and appropriate models are KERRBB (\citealt{Lietal}) and AGNSED (KD18)\footnote{Both models are implemented in XSPEC (\citealt{ArnaudXPS}).}: the former describes a geometrically thin AD including all relativistic effects (e.g., frame dragging, returning radiation, light bending, gravitational redshift); the latter also takes into account the presence of an X-ray corona above the disk (even though relativistic effects such as the ones implemented in KERRBB are not fully included), thus requiring X-ray data to constrain the model parameters.\footnote{Relativistic effects due to large spin values modify the AD emission shape with respect to the "standard" one (e.g., \citealt{SS}; \citealt{NovTho}): AGNSED does not include all such effects possibly affecting the predicted disk and corona emissions.} In this framework, given the large uncertainties on the physical and geometrical properties of the X-ray corona (see Sect. \ref{AGNSED_3} and Appendix \ref{APP:AGNSED} for a discussion about AGNSED) and the importance of relativistic effects, we adopted KERRBB to estimate the SMBH masses using the analytical expressions found by C18. We then considered the uncertainties of the model by comparing BH masses inferred with KERRBB with those from AGNSED for two sources with good data coverage in the X band. Significant uncertainties on the results also arise from observational issues such as absorption by dust. These possible uncertainties are discussed below along with their weight on our estimates.

The thin-disk approximation breaks down for Eddington ratios $\gsim 0.3$ (e.g., \citealt{LaoNet}) and for this reason AD models in the so-called "slim" and "thick" regimes should be considered. One of these is the relativistic model SLIMBH (\citealt{Abretal}; \citealt{Sad09}; \citealt{SadwAbra09}; \citealt{SadwAbra}): as shown by C19, for a given set of data, the difference in the BH masses derived from KERRBB and SLIMBH is less than a factor of $\sim 1.2$ (see the discussion in Appendix \ref{AP:AGNSED}); therefore, for the purpose of this work, we use KERRBB as a good enough approximation of the AD emission.

As already mentioned, we intend to compare the BH masses estimated from the AD fitting with those from the RM technique. This choice is due to two factors: RM is rather direct in the sense that it is not based on calibrated statistical relations but on individual measurements; and there is a significant (and still growing) number of sources with RM BH mass estimates. We selected our sample from the AGN Black Hole Mass Database (\citealt{BenKat})\footnote{See also http://www.astro.gsu.edu/AGNmass/.}, a compilation of published spectroscopic RM studies of AGNs, choosing sources with clear evidence of the Big Blue Bump.

The paper is structured as follows: in Sects. \ref{sec:2} and \ref{Fitting}, the AGN sample and the AD fitting procedure are described in detail. Section \ref{Modifiche} illustrates possible issues related to absorption and to uncertainties of the model. The BH masses estimated through the AD fitting are reported and compared with those from RM and SE in Sect. \ref{sec:results}; we also discuss the possibility of estimating the BH spin and the comparison of our results with the ones inferred using AGNSED. In Sect. \ref{sec:dis} we discuss the results and present our conclusions. In the Appendices (A, B and C), we present the basic equations used for the fits with KERRBB (A), the comparison between KERRBB and other AD models (B), the list of the sources and their fits, inferred masses and Eddington ratios (C). 

In this work, we adopt a flat $\Lambda$CDM cosmology with $H_0=67.4$ km s$^{-1}$ Mpc$^{-1}$ and $\Omega_{\rm M}=0.315$ (Planck 2018 Results).

\section{Sample and data selection} \label{sec:2}

Here we define the AGN sample and describe the AD fitting procedure, illustrating the possible issues related to this approach. 

For all the sources of the AGN Black Hole Mass Database, we searched for the available and the most recent spectroscopic data from the near-infrared (NIR) to the far-ultraviolet (FUV) band from the public archives and literature (see Tab. \ref{ap:alltabresres_data} and Appendix \ref{AP:C}).\footnote{We also collected some photometric data (GALEX, Vizier, NED), not taken into account in the fitting procedure because (i) they might be contaminated by emission lines or by some kind of absorption and (ii) their statistical weight in the fit is negligible.} Among all sources, we then selected 28 ($z<0.3$) with (i) a clear UV bump determined as a power-law continuum $F_{\lambda} \propto \lambda^{\alpha}$ with a negative slope in the rest-frame wavelength range $3000 - 5000$ \AA \footnote{Only two sources (PG1247+267, S50836+71) are at $z \sim 2$. For them, spectroscopic data cover only the spectrum peak position.}, (ii) wide spectroscopic coverage especially around the spectral peak, and (iii) limited variability for non-simultaneous spectra (see below). For some sources, the NIR-optical SED shows contamination by the host galaxy whose emission was taken into account in the SED modeling (see following section). For each source, spectroscopic data were corrected for Galactic extinction using the \citet{Cardelli} reddening law and $E[B-V]$ from the map of \citet{Schle} with $R_{\rm V} = 3.1$.

A serious issue concerns non-simultaneous spectroscopic data that could affect the normalization of different spectra. When data did not show the same normalization, we calibrated the different data sets by adopting the following procedure: first, we considered spectroscopic data in the UV wavelength range where the peak of the AD emission should be located; then we calibrated the available FUV and optical data by matching the flux in the common wavelength range assuming that the spectral shape does not change with flux variation (see also \citealt{Shang05}). The same calibration was performed on IR data (when present). In any case, the maximum mismatch amounts to $\sim 0.1$ dex in flux (leading to an uncertainty on the derived BH mass at most by a factor of $\sim 0.05$ dex). 

Finally, spectroscopic data (especially FUSE and HUT data) were smoothed by averaging the flux in fixed wavelength bins in order to have a clearer representation of the overall emission. This latter process could have an effect on the fitting procedure and on the inferred model parameters (see following section).

\section{Fitting procedure}\label{Fitting}

\begin{figure}
\centering
\hskip -0.2 cm
\includegraphics[width=0.49\textwidth]{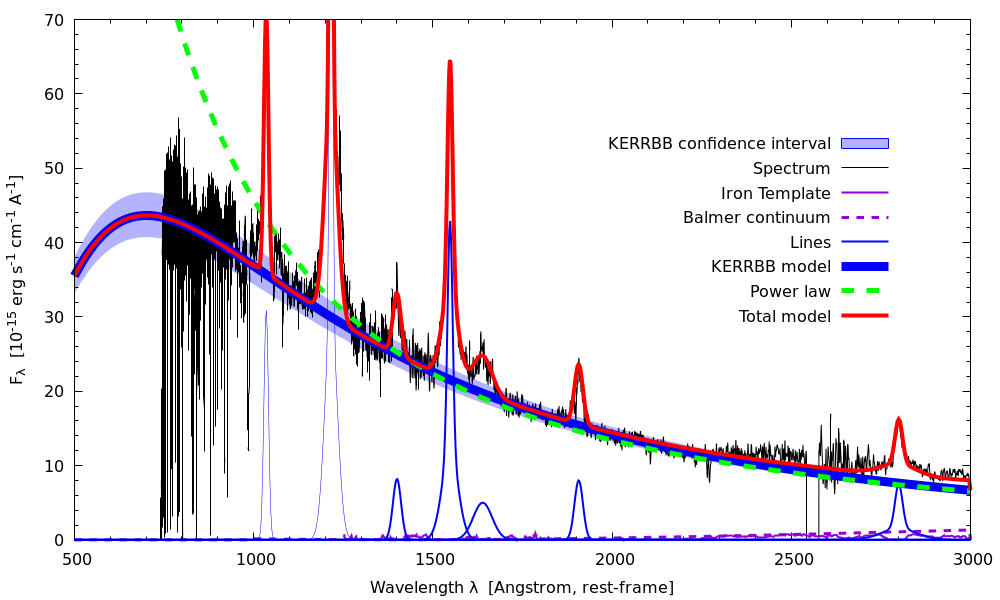}
\caption{Example of a fit of the composite FUSE - HST - KPNO spectrum of the source PG0953+414 (from \citealt{Shang05}). The modeling performed with KERRBB (thick blue line) describes the AGN continuum, the iron complex (purple line), the Balmer continuum (purple dashed line), and some prominent emission lines (MgII, CIII, CIV, SiIV, Ly$\alpha$) described by a simple Gaussian profile (blue lines). The red line represents the sum of all these components. The green dashed line is a standard power-law continuum (slope $\alpha = -1.77$): for $\lambda > 1300\ \text{\AA}$, the KERRBB model overlaps the power-law rather well within the average confidence interval ($\sim 0.05$ dex, blue shaded area) and is not affected significantly by the presence of lines or other spectral features. Abundant interstellar absorption lines at $\lambda < 1000$ \AA have a negligible effect on the determination of the spectral peak emission (see Sect. \ref{Modifiche}).} 
\label{fit_spettro}
\end{figure}

We adopted the relativistic thin AD model KERRBB for the fitting procedure. We used GNUPLOT (non-linear least-squares Marquardt-Levenberg algorithm) to fit the rest-frame spectrum ($\lambda - F_{\lambda}$, Fig. \ref{fit_spettro}) with the KERRBB model to describe the AGN continuum, adding the iron complex (e.g., \citealt{VesterWi}), some prominent emission lines (MgII, CIII, CIV, SIV and Ly$\alpha$) modeled with a simple Gaussian profile, a Balmer continuum (e.g., \citealt{Derosa}) and the template for the host-galaxy emission (\citealt{Manuc}) which can contaminate the nuclear spectrum in the NIR-optical bands.
	
As a comparison, we also performed a standard fit using a power-law to describe the AGN continuum: in Fig. \ref{fit_spettro} we show, as an example, the results of both fits. It is clear that the two continua overlap well for $\lambda > 1300\ \text{\AA}$ while the AD model turns over around the spectral peak. Even if this latter is not covered by data, it can be constrained using the curvature of the spectrum at smaller frequencies. Spectral features, such as strong emission or absorption lines, have no drastic effects on the overall KERRBB fit, even at short wavelengths where the spectrum peak is located (see following section).\footnote{As an additional test, we used two Gaussians to fit the broad base of the most prominent lines: we still have no drastic effects on the AGN continuum which remains inside the confidence interval of $\sim 0.05$ dex.} 

Each fit constrains the spectrum peak position which is essential to infer the BH mass (see Appendix \ref{AP_B}): in the $\nu - \nu L_{\nu}$ representation, both the peak frequency $\nu_{\rm p}$ and luminosity $\nu_{\rm p} L_{\nu_{\rm p}}$ have a mean uncertainty of $\sim 0.05$ dex (represented with a blue shaded area in Fig. \ref{fit_spettro} and in all the Figures reporting the spectral fits, Figs. \ref{SED10} - \ref{SED39}). This confidence interval translates to an average uncertainty of $\sim 0.1$ dex on the BH mass (for a fixed BH spin and viewing angle - see Sect. \ref{5.1}).

\section{Caveats: data uncertainties}\label{Modifiche}

In this section we discuss the possible effects of absorption by gas or dust on the overall AD spectrum shape.

First, for some ground-based telescopes, the available spectrum can be subjected to absorption due to sky regions with low transparency. Even if these regions are subtracted, our best fit does not change significantly and remains inside the average confidence interval ($\sim 0.05$ dex).	

Spectra show some absorption features caused by the interstellar medium, especially at frequencies Log $\nu /{\rm Hz} \gsim 15.4$: if these blended lines are smoothed or not subtracted from the spectrum appropriately, the AGN continuum can be underestimated, leading to an incorrect evaluation of the spectral peak position (i.e., shifting $\nu_{\rm p}$ to smaller values and leading to an overestimation of the BH mass). In order to understand if these spectral features have a relevant effect on our results, we performed the same fitting procedure described in the previous section, choosing only the frequency range Log $\nu /{\rm Hz} \lsim 15.4$; even if the spectral range is reduced, the curvature of the spectrum at smaller frequencies can still be used to constrain the peak position. We find that the new peak frequencies are on average larger than the previous ones (inferred considering the whole available spectral range) but consistent with them within a range of $\sim 0.05$ dex, while the luminosities are similar. The inferred new BH masses are inside the average confidence interval of $\sim 0.1$ dex (defined by the spectrum peak uncertainty).
	
\begin{figure*}
\centering
\hskip -0.2 cm
\includegraphics[width=0.48\textwidth]{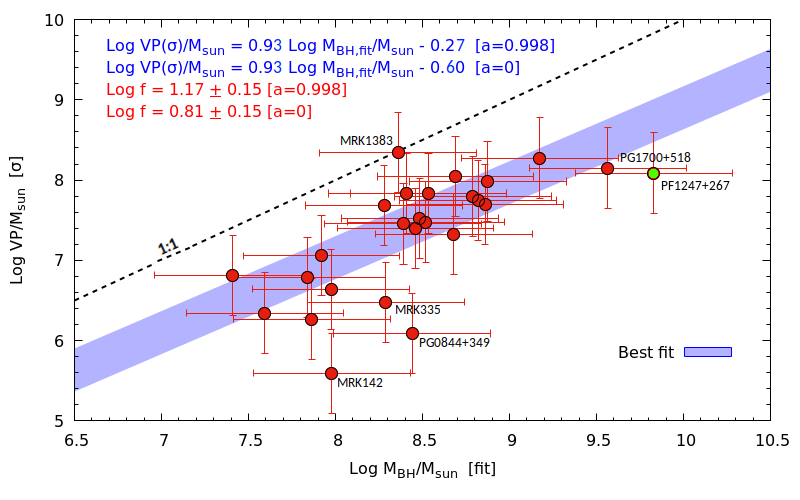}\includegraphics[width=0.48\textwidth]{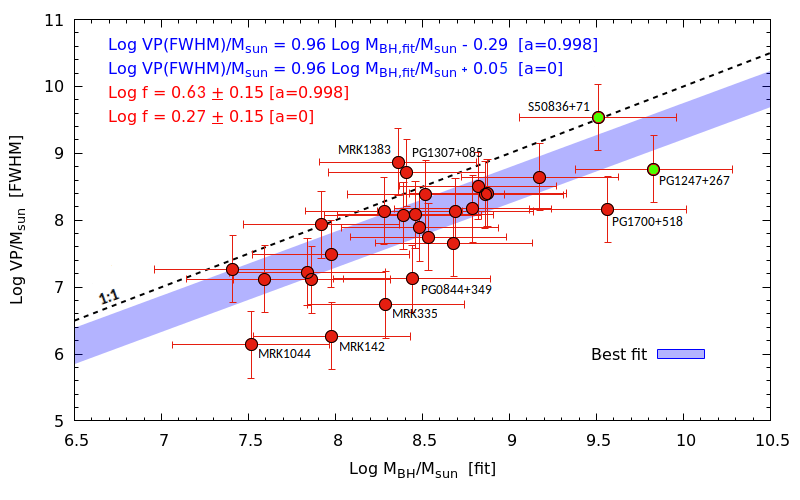} 
\caption{Comparison between KERRBB BH mass estimates $M_{\rm BH, fit}$ (inferred from the SED fitting procedure; each dot corresponds to the mean value computed using the extreme estimates given by the uncertainties) and the VPs calculated using the H$\beta$ velocity dispersion $\sigma_{\rm line}$ (left panel) and the FWHM (right panel). We averaged all the VPs computed using data from different authors (see Tables \ref{tab:res} - \ref{tab:res3}; PG1247+267 and S50836+71 are marked with green dots). The blue shaded area corresponds to the set of best fits between $M_{\rm BH,Fit}$ and VP after considering all possible spin values between 0 and 0.998 (equations in blue on the plot are the two extreme cases; the uncertainty on the slope is $\sim 20 \%$). Assuming that Log $M_{\rm BH,fit} = {\rm Log VP} + {\rm Log} f$, we found the scale factor $f$, labeled on the plot in red for the two extreme spin values. Uncertainty bars in both plots are $\sim 0.45$ dex and $\sim 0.5$ dex for $M_{\rm BH, fit}$ and VPs respectively, and the black dashed line is the 1:1 line.} 
\label{MASS_RM_fit}
\end{figure*}

The intergalactic medium (IGM) could also modify the spectrum shape (especially for high-redshift sources) at frequencies Log $\nu /{\rm Hz} \gsim 15.4$. For the redshift range spanned by our sample the correction from IGM absorption is negligible (see \citet{Madau}; \citet{HaaMad}; \citealt{Castietal}) except for the two high-redshift sources, PG1247+267 and S50836+71. For these latter, we performed this correction and showed the results in Figs. \ref{SED34} - \ref{SED39}, respectively.

An important effect that could modify the spectral UV shape concerns dust absorption: if present, this could lead to an incorrect BH mass estimate. The absorption could be caused by the dusty torus surrounding the AD, or dust in the host galaxy interstellar medium (ISM).\footnote{Recent works (e.g. \citealt{Left}) show that $\sim 50 - 80 \%$ of the MIR emission originates primarily from polar regions instead of from an equatorial dust distribution. This can have an effect on the observed disk luminosity which would be dimmer with respect the intrinsic one since part of its radiation is intercepted by the polar dust.} Given that our sample is composed of Type 1 QSOs (${\rm FWHM} > 1000$ km/s for the most prominent lines; e.g. \citealt{Anto}) we do not expect any (strong) absorption from the dusty torus, which is assumed to have an average opening angle of $\sim 45^{\circ}$. For this reason, in order to infer the BH mass, we assumed that each source was observed with a viewing angle $\theta_{\rm v} \leq 45^{\circ}$ (see Appendix for results). 

However, we checked the goodness of our fit by considering the possible intrinsic reddening by the host galaxy ISM. To do so, we followed the work by \citet{Baron}, who found an analytical expression to infer the amount of absorption (in terms of $E[B-V]$) as a function of the rest frame spectrum slope $\alpha_{\nu}$ in the wavelength range 3000 - 5100$\ \text{\AA}$. We find that for the majority of the sample, the correction is small ($E[B-V] < 0.05$ mag) and could lead to a decrease of the BH mass by a factor $\lsim 0.1$ dex. The extinction found is consistent with what is thought to be the average value for AGN ($E[B-V] \sim 0.05 -0.1$ mag; e.g., \citealt{Kora}).

For a more complete analysis of the possible UV dust absorption, for each source we used the extinction law of \citet{Czetal} and assumed that the slope of the corrected, de-reddened spectrum at wavelength $< 2000\ \text{\AA}$ had to be softer than the theoretical value $F_{\nu} \propto \nu^{1/3}$. In this way, we find an upper limit for the correction (on average $E[B-V] \sim 0.20$ mag) that leads to a decrease of the BH mass obtained through the SED fitting procedure at most by a factor of $\sim 0.3$ dex (since the spectrum peak position changes due to the correction). 

As explained above, we do not expect such a strong UV absorption because we are dealing with Type 1 QSOs. Moreover, for what concerns the correction found by \citet{Baron} regarding ISM dust absorption, possible deviations from the average continuum slope could be caused by other factors connected to the BH physics, such as the BH mass, the accretion rate, the spin, and the system orientation (e.g., \citealt{Hub2000}; \citealt{DavLao}). Therefore, we did not consider any correction from dust absorption, and are confident that our results are inside the estimated BH mass confidence interval (even if the intrinsic extinction is taken into account).

\section{Results} \label{sec:results}

In this section, we show the results coming from the SED fitting procedure. We used the analytical expressions found by C18 in order to infer the BH mass and the Eddington ratio (reported in Tables \ref{tab:res} - \ref{tab:res3}) using KERRBB, in the spin range $0 \leq a \leq 0.998$ and for a viewing angle $\theta_{\rm v} \leq 45^{\circ}$. As shown by the authors, the space of KERRBB solutions is degenerate and by using different parameters ($M$, $a$ and accretion rate) appropriately, it is possible to describe the same set of data.

We compare KERRBB BH masses $M_{\rm BH,Fit}$ to the ones obtained through RM and from SE equations: we fit the relation between our estimates linearly, that is, the virial product (VP; computed with RM results) and SE estimates (i.e., Log $M_{\rm BH,Fit} = {\rm Log} M[{\rm VP\ _{or}\ SE}] + {\rm Log} f$). The VP is the base of all virial-based mass measurements:
\begin{equation}\label{virialp}
	M_{\rm BH} = f_{\rm BLR} \underbrace{\frac{c \tau_{LT}\ \sigma_{\rm line}^2}{G}}_{= {\rm VP}},
\end{equation} 

\noindent where $\tau_{\rm LT}$ is the light-travel time (i.e., the time related to the emission-line response delayed with respect to changes in the continuum) and $\sigma_{\rm line}$ is the line velocity dispersion (or the line FWHM; e.g., \citealt{Ho}; \citealt{Wanetal}). 
The factor $f$ corresponds to the geometrical factor $f_{\rm BLR}$ when the comparison is performed with the VP, while it is merely a multiplicative factor in the case of SE measurements, for which $f_{\rm BLR}$ has already be chosen in literature as an average value ($\sim 1$, \citealt{VesterPeter}). Many authors have calibrated the geometrical factor by comparing BH masses obtained with different approaches (see \citealt{BenKat}, and references therein). \footnote{See \citet{Li18} for a list of $f_{\rm BLR}$ values found in literature.} Some authors use $f_{\rm BLR}=3$ ($3/4$), if the VP is computed using $\sigma_{\rm line}$ (FWHM), considering a spherical distribution of BLR clouds in randomly orientated orbits (\citealt{Net}; \citealt{Wanetal}; \citealt{Kaspietal}).\footnote{The line velocity dispersion is often identified either as the line FWHM or as the $\sigma$ of the Gaussian profile used to fit the emission line.} The RM technique estimates have a systematic uncertainty of $\sim 0.5$ dex (as the SE one,  \citealt{VesterOs}).

\subsection{Black hole mass uncertainty}\label{5.1}

The total uncertainty on the KERRBB BH mass inferred from the SED fitting procedure is $\sim 0.45$ dex (comparable to the systematic uncertainties on the RM and SE estimates: $\sim 0.5$ dex). This uncertainty has to be considered as a confidence interval in which the BH mass inferred with KERRBB lies and is connected to different quantities involved in the fitting procedure, namely the BH Spin, in the range $0 \leq a \leq 0.998$; the viewing angle of the system, in the range $0^{\circ} \leq \theta_{\rm v} \leq 45^{\circ}$; and the uncertainty on the spectral peak frequency and luminosity.

Assuming that there is no dust absorption, the BH mass changes by $\sim 0.5$ dex going from $a=0$ to $a=0.998$ (for a fixed viewing angle), and by $\sim 0.2$ dex going from $\theta_{\rm v} = 0^{\circ}$ to $\theta_{\rm v} = 45^{\circ}$ (for a fixed spin). Taking as a reference value the arithmetic mean of the BH mass in both the spin and $\theta_{\rm v}$ ranges, the overall uncertainty is $\sim 0.35$ dex. Moreover, the confidence interval on the spectral peak position ($\sim 0.05$ dex) leads to an additional uncertainty of $\sim 0.1$ dex on the BH mass estimate. However, if the spectral peak is well constrained (and/or the viewing angle is known), the mean uncertainty on the AD BH mass estimates can be $\lsim 0.3$ dex.

\subsection{Black hole mass comparison}

\begin{figure}
\centering
\hskip -0.2 cm
\includegraphics[width=0.49\textwidth]{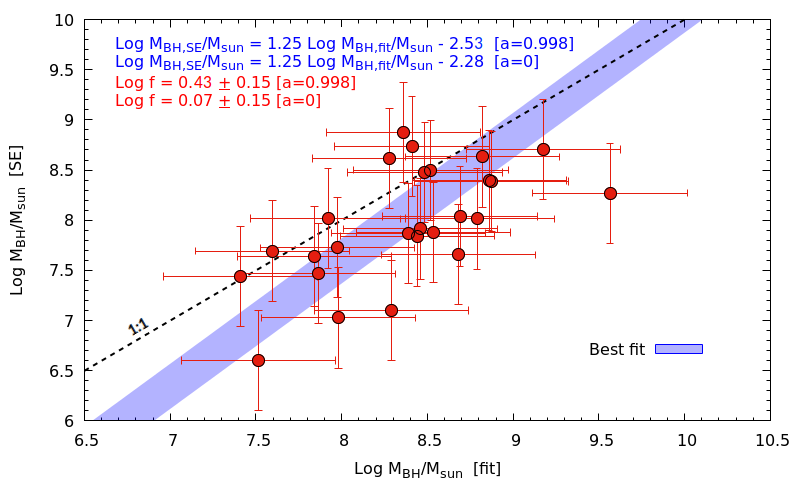}
\caption{Comparison between the KERRBB BH mass estimates $M_{\rm BH, fit}$ (inferred from the SED fitting procedure) and the SE BH masses $M_{\rm BH, SE}$ computed using the equation of \citet{VesterPeter} and the H$\beta$ line (we excluded the two high-redshift sources, PG1247+267 and S50836+71, for which we only have information on the CIV line; see Table \ref{tab:res}). The blue shaded area, the dashed black line, the reported labels, and the uncertainty bars are the same as in Fig. \ref{MASS_RM_fit}.} 
\label{MASS_SEnew_fit}
\end{figure}

Figure \ref{MASS_RM_fit} shows the comparison between our BH mass estimates inferred from the SED fitting procedure with KERRBB and the VPs computed using the H$\beta$ velocity dispersion $\sigma_{\rm line}$ (left panel) and the FWHM (right panel; as we used the CIV line to compute the VPs, we excluded PG1247+267 and S5 0836+71 from the fit for consistency). The comparison between the VP computed using the velocity dispersion and the FWHM is shown because several authors claimed that the ratio FWHM/$\sigma_{\rm line}$ is not necessarily a constant (e.g., \citealt{Collin}, \citealt{Peterson11}): for our sample, we find FWHM/$\sigma_{\rm lines} \sim 2$ with a large dispersion ($\sim 0.5$ dex). Instead, Fig. \ref{MASS_SEnew_fit} shows the comparison between our results and the BH masses computed using the SE relations of \citet{VesterPeter}. 

From the analysis of those results, we find that both the VPs and the SE estimates are systematically smaller than the KERRBB estimates by a factor $f$ depending on the BH spin. For VP$(\sigma_{\rm line})$ and VP(FWHM), assuming a BH spin $a= 0\ (0.998)$, we find Log $f = 0.81\ (1.17) \pm 0.15$ and Log $f = 0.27\ (0.63) \pm 0.15$, respectively. As in this case $f$ is identified as the geometrical factor $f_{\rm BLR}$ (beginning of Sect. \ref{sec:results}), the range we find by using VP($\sigma_{\rm line}$) is consistent for example with $f_{\rm BLR} = 5.5 \pm 1.8$ (\citealt{Onkenetal}; see also \citealt{Li18} for other reference values). In both cases, the compatibility is more compelling if, on average, BH spins are assumed to be small. For SE estimates, we find Log $f = 0.07\ (0.43) \pm 0.15$, assuming a BH spin $a= 0\ (0.998)$, partially consistent with the recent work by \citet{Marcule}. This result is similar to the one found for VP(FWHM): this is due to the use of a geometrical factor $f_{\rm BLR} \sim 1$ and FWHM measurements inside virial equations of \citet{VesterPeter}. Therefore SE and VP(FWHM) are almost the same (within uncertainties).

\begin{figure*}
\centering
\hskip -0.2 cm
\includegraphics[width=0.8\textwidth]{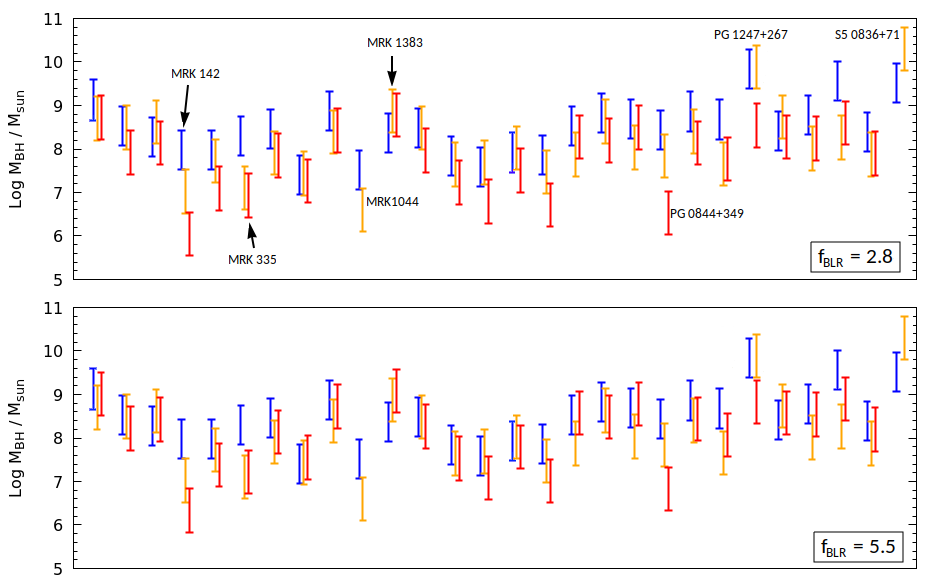}
\caption{Comparison between the BH masses computed from the SED fitting procedure with KERRBB (blue) and the SE (orange) and RM estimates (red, inferred with $\sigma_{\rm line}$ and a geometrical factor $f_{\rm BLR} = 2.8 - 5.5$, \citealt{Graham} top panel and \citealt{Onkenetal} bottom panel). The average uncertainty for all the measurements is $\sim 0.5$ dex (the results related to each source are plotted in the same order as listed in the Tables of Sect. \ref{AP:C}).} 
\label{MASS_SEVP_comp_fit_ecc}
\end{figure*}

From Fig. \ref{MASS_RM_fit} (left panel), it is clear that KERRBB mass estimates are systematically larger than the corresponding VPs: from the fit, a better compatibility is reached if a scale factor of the order of $\sim 10$ is assumed, in agreement with the recent paper by \citet{Pozo}. Nonetheless, our results are still compatible with those estimated using RM if, for these latter, a geometrical factor of less than ten is considered. Figure \ref{MASS_SEVP_comp_fit_ecc} shows our KERRBB BH estimates compared with the ones from SE and RM (computed using the velocity dispersion $\sigma_{\rm line}$ and different geometrical factors $f_{\rm BLR}=2.8 - 5.5$, from \citealt{Graham} and \citealt{Onkenetal}, respectively): about $70 \%$ of the sources show a good compatibility between the three results within uncertainties ($\sim 0.5$ dex, on average) favoring smaller KERRBB masses and therefore low spin values. See Sect. \ref{AGNSED_3} and Appendix \ref{AP:AGNSED} for a comparison between KERRBB results and those inferred with other models..

\subsection{Spin estimate}\label{KBB_spin}

Given the large uncertainties involved in each method to infer the BH mass, an estimate of the BH spin is still a hard task. In principle, the comparison of two or more independent BH masses with relativistic AD model results could be an alternative way to constrain the BH spin (e.g., see C19). 

For the majority of the sources, the comparison showed that small spin values are favored. Instead, for a couple of sources (MRK 1383, S50836+71), the comparison between RM, SE, and KERRBB results (within uncertainties) led to the conclusion that the BH spin must be high (we found a lower limit $a>0.6$). However, as mentioned above, because uncertainties on the fitting procedure, on the parameters of the AD, and on the RM measurements (as well as other methods) are still large, we suggest caution be taken when considering a BH spin estimate based on the comparison of different BH masses.

\subsection{X-ray corona above the disk: modifications in the BH mass estimates}\label{AGNSED_3}

Estimations of the BH mass using the SED fitting procedure could be affected by the presence of a hot Corona above the disk: this structure is thought to be compact (e.g., \citealt{Miniu}; \citealt{Done}; \citealt{ReiMil}; \citealt{Sazo}; \citealt{LuRi}) and responsible for the emission in the X band and for the soft X-ray excess observed in many AGNs (e.g., \citealt{Kora}). In principle, if this structure scatters a fraction of the disk radiation, the observed disk luminosity could be dimmer than the intrinsic one, leading to an incorrect mass estimate (i.e., the spectrum peak position could be different from the intrinsic one). 

In order to check this possibility, we compared our results with those of the relativistic model AGNSED (KD18), which also takes into account also the presence of an X-ray corona above the disk in a self-consistent way. It is important to note that, contrary to KERRBB, AGNSED does not include relativistic effects, such as light bending and gravitational redshift, which may have a significant effect on both the AD and the corona emissions; however, for small values of the BH spin and $\theta_{\rm v}$, those effects should have a minor weight on the results. 

For the fitting procedure, we used the sources NGC5548 and MRK509 (see Appendix \ref{AP:AGNSED} for details) also studied by KD18. We found that, as for KERRBB, the modeling with AGNSED of the optical-UV-X SED is degenerate: by changing the model parameters (i.e., BH mass, spin, Eddington ratio, corona size) appropriately, it is possible to reproduce the same set of data (for some reference values, see Fig. \ref{AGNSED}).

For NGC5548, results from both models are compatible while for MRK509, KERRBB BH masses are larger than the AGNSED ones by a factor $\lsim 0.2$ dex (Fig. \ref{AGNSED22} right panel). We argue that for NGC5548, the absence of relativistic effects in AGNSED is balanced out by modeling a large corona above the disk, leading to the same BH mass. Instead, for MRK509, the smaller X-ray corona leads to different results because it does not compensate the differences between the two models. 

In general, KERRBB observed disk luminosity is dimmer than the intrinsic one because of a compact corona located in the inner region of the disk, the presence of which modifies the AD emission and leads to an overestimated BH mass. By correcting the KERRBB disk lumminosity by a factor $\lsim 0.2$ dex, results from the two models related to MRK509 show better compatibility (see Appendix \ref{AP:AGNSED} for more details about this correction).

The partial compatibility between KERRBB and AGNSED shows that BH masses estimated though the SED fitting procedure clearly depend on the adopted model and their physical background. Despite these findings, KERRBB results for both sources are still compatible with SE and RM ones within uncertainties (Fig. \ref{MASS_SEVP_comp_fit_ecc}). Moreover, results presented by KD18 are compatible with ours for what concerns the BH spin: taking their finding for the BH masses as reference values, both their results and ours favor low values of $a$.

\section{Discussion and conclusions} \label{sec:dis}

We used the relativistic thin AD model KERRBB (\citealt{Lietal}) to infer the BH masses of 28 sources from the AGN Black Hole Mass Database (\citealt{BenKat}). These sources have a BH mass estimate from the RM studies and show clear evidence in the UV band of the so-called Big Blue Bump produced by the radiation coming from an AD around a SMBH. Since we did not have information about the viewing angle for the majority of them, we assumed that the Type 1 QSOs were observed with $\theta_{\rm v} \leq 45^{\circ}$ in order to avoid the absorption from the dusty torus (assumed to have an average opening angle of $\sim 45^{\circ}$). Our results ca be summarized as follows:
\begin{itemize}
	\item The majority of the sources of our sample show a good match between data and the modeling with KERRBB. The modeling led to a relatively good estimate of the spectrum peak position (i.e., peak frequency and luminosity; see Appendix \ref{AP_B}) with a small uncertainty ($\sim 0.05$ dex on average).
	\item The total uncertainty related to the AD BH mass estimates is $\sim 0.45$ dex, which is connected to the unknown BH spin (in the range $0 \leq a \leq 0.998$), the viewing angle ($\theta_{\rm v} \leq 45^{\circ}$), and the uncertainty on the peak position from the fitting procedure. If the quality of the data is high and either the spectral peak or the viewing angle of the system are well constrained (with an uncertainty of less then $\sim 10 \%$), the mean uncertainty is reduced to $\lsim 0.3$ dex, which is smaller than the systematic uncertainties on both the RM and SE estimates ($\sim 0.5$ dex, e.g., \citealt{VesterOs});
	\item Our BH mass estimates with KERRBB are systematically larger than the corresponding VPs (computed using both the velocity dispersion $\sigma_{\rm line}$ and the FWHM). We computed the difference through a scale factor $f$ and found Log $f \lsim 1.2$ (depending on the BH spin - Fig. \ref{MASS_RM_fit}), which is also in agreement with the recent paper by \citet{Pozo}. A similar result is found from the comparison between our results and BH masses inferred with SE equations (Log $f \lsim 0.5$, Fig. \ref{MASS_SEnew_fit}).	
	
	\item Despite these findings, assuming a geometrical factor in the range $f_{\rm BLR} = 2.8 - 5.5$ (from \citealt{Graham} and \citealt{Onkenetal}, respectively), we find a compatibility between RM, SE, and KERRBB results for $\sim 70\%$ (Fig. \ref{MASS_SEVP_comp_fit_ecc}).
	
	\item In principle, a comparison between independent BH mass estimates (e.g., RM, SE) and the ones coming from the SED fitting procedure with a relativistic AD model could lead to an estimate of BH spin (e.g., C19): for a couple of sources we find a lower limit ($a>0.6$) but, for the majority of the sample, low spin values are favored. These results must be considered with caution because the uncertainties on the different parameters involved in the fitting procedure and RM (or SE) measurements are still large. 
	
\begin{figure}
\centering
\hskip -0.2 cm
\includegraphics[width=0.495\textwidth]{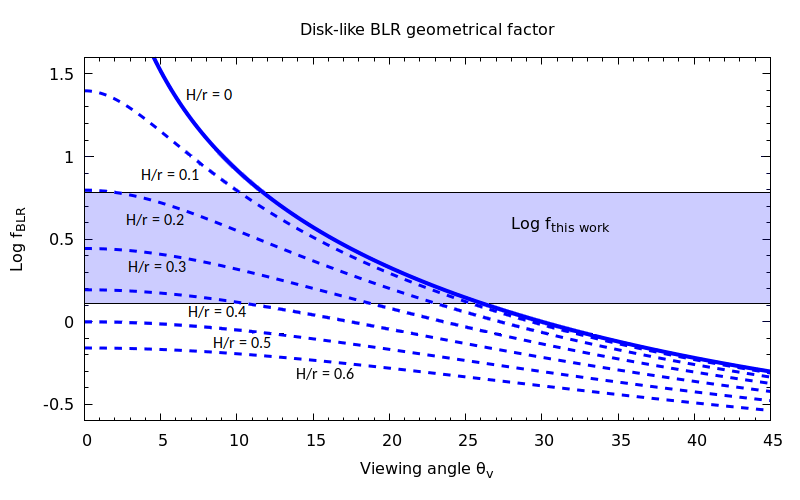}
\caption{Geometrical factor $f_{\rm BLR}$ computed using Eq. \ref{eq:fblr} (e.g., \citealt{Collin}; \citealt{Decarli}) as a function of the viewing angle for different BLR thickness $H/r$. The blue shaded area is the scale factor range found in this work (related to VPs computed using the FWHM; see Fig. \ref{MASS_RM_fit}, right panel). The comparison leads to a range of $\theta_{\rm v}$ and $H/r$ consistent with the work of \citet{Maj16}. }
\label{FBLR}
\end{figure}

	\item Assuming that the BH masses estimated through the AD fitting procedure are correct, the geometrical factor must be large (Fig. \ref{MASS_RM_fit}) in order to compensate for the difference with the corresponding VPs. Assuming a disk-like BLR (e.g., \citealt{Collin}; \citealt{Decarli}), the geometrical factor related to the VP (computed using the FWHM) is:
\begin{equation}\label{eq:fblr}
	f_{\rm BLR} = \frac{1}{4\ \Bigl[ (\sin \theta_{\rm v})^2 + (H/r)^2  \Bigl]}
\end{equation}
	
	 where $H/r$ is the height-to-radius ratio (i.e., thickness) of the BLR. The range of scale factors found in this work is consistent with a BLR seen with a viewing angle $<30^{\circ}$ (consistent with Type 1 QSOs) and a thickness $H/r < 0.5$ (consistent with the results of \citealt{Maj16}; see Fig. \ref{FBLR}).

	\item Black hole mass measurements through the SED fitting procedure could be affected by the presence of a hot X-ray corona above the AD. We checked this possibility by comparing our KERRBB results with those from the relativistic model AGNSED (KD18) that accounts also for the X-ray emission in a self-consistent way. We used two sources, NGC5548 and MRK509, and find that for NGC5548, results from both models are compatible while for MRK509, KERRBB BH masses are larger by a factor $\lsim 0.2$ dex with respect to those inferred with AGNSED. We argue that the possible presence of an X-ray corona above the disk modifies the emission of this latter leading to an overestimated BH mass; we corrected the emission following the simple prescription described in Appendix \ref{AP:AGNSED} and reaching a relatively good compatibility between the two results. However, despite this correction, if uncertainties on RM and SED fitting estimates are taken into account, both our results and those of KD18 favor small BH spin values.
 
	\item The comparison between all those results suggests that the systematic discrepancy between KERRBB and RM (or SE) masses (i.e., the large value of the factor $f$ - see Fig. \ref{MASS_RM_fit}) could possibly be related to the choice of the AD model adopted in the SED fitting procedure. However, the partial compatibility between KERRBB and AGNSED leads to the conclusion that results are strictly connected to the geometry of the X-ray corona and to the relativistic AD radiation pattern, which are not both included in the models used here. Furthermore, a comparison between different models is necessary in order to understand their differences and to improve them.
\end{itemize}

Despite the large uncertainties involved in the fitting procedure and RM (or SE) measurements, KERRBB showed a good agreement with data, strengthening the choice of AD models as an alternative method both to described the observed SEDs and to infer the mass of SMBHs. For what concerns these latter arguments, the choice of the AD model is crucial even though we found similar results using different models. Nonetheless, although few attempts are discussed in this work, the uncertainties involved in these kinds of measurements are still too large to have precise information on the BH accretion and spin from the comparison between different methods. A larger sample of sources with RM measurements and a clear prominent AD emission are necessary to strengthen these findings, to improve AD models for what concerns the modeling of the observed SEDs, and possibly to obtain more information on the BH accretion and rotation.

%-------------------------------------------

\begin{acknowledgements}
 
We thank the anonymous referee for her/his constructive comments and suggestions, useful to improve the manuscript.

\end{acknowledgements}

%-------------------------------------------

\let\itshape\upshape

%-------------------------------------------

\appendix

\section{Fit, BH mass, and Eddington ratio}\label{AP:C}

We show the fit of the individual SED performed by adapting the KERRBB model to the rest-frame spectrum (data and results are reported in Tables \ref{ap:alltabresres_data} - \ref{tab:res3}). When necessary, a host-galaxy template (from \citealt{Manuc}) is added to the KERRBB model in order to obtain a better fit in the frequency range Log $\nu / {\rm Hz} < 14.8$. For a few sources, we used IUE data instead of the most recent HST ones because the former covers a wider wavelength range. Red dots are archival photometric data (GALEX, Vizier, NED) not used in the fitting procedure because they might be contaminated by emission lines or some kind of absorption. Along with the KERRBB best fit, we show a blue shaded area ($\sim 0.05$ dex) that defines a confidence interval for the spectrum peak position.

For two sources (PG0844+349, PG1211+143), we corrected the spectrum from possible dust absorption (following \citealt{Czetal}) in order to have a better compatibility between the data and KERRBB: for PG 0844+349, a KERRBB + host-galaxy modeling cannot describe the overall spectrum and for this reason we showed a better compatibility by correcting it from dust absorption; instead, for PG1211+143, correcting the spectrum from dust leads to a satisfactory fit even without including the host-galaxy emission. Despite this correction, BH masses do not change drastically ($\lsim 0.1$ dex). For the high-redshift sources (PG1247+267, S50836+71, $z \sim 2$), we tried to correct the spectrum (and photometric data) from IGM absorption, following \citet{Madau}, \citet{HaaMad} and \citet{Castietal} showing the corrected results in the corresponding plots.

\begin{figure*}[b]
\centering
\hskip -0.2 cm
\includegraphics[width=0.505\textwidth]{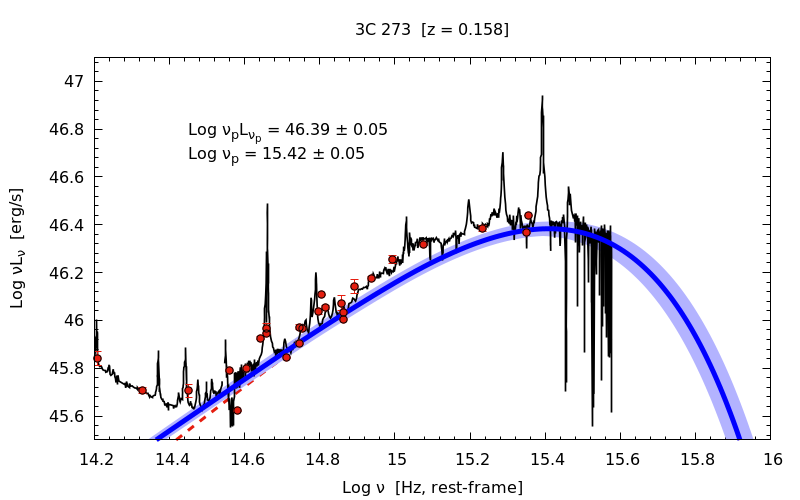}\includegraphics[width=0.505\textwidth]{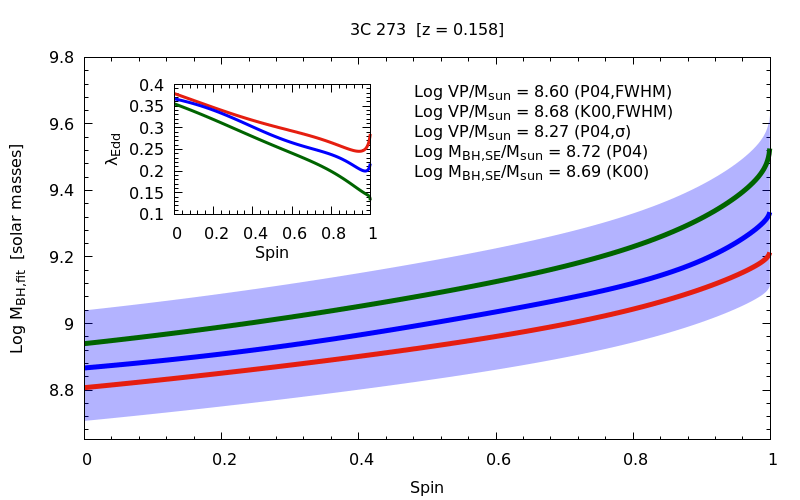}
\caption{Other recent BH mass estimates are from \citet{Sturm} (Log $M/M_{\odot} = 8.41^{+0.16}_{-0.23}$) and \citet{Zhangetal19} (Log ${\rm VP} ({\rm FWHM}) /M_{\odot} = 8.50^{+0.04}_{-0.06}$). On each of the following plots, we label different BH mass estimates (VPs computed using the $\sigma_{\rm line}$ and FWHM, and SE estimates computed using the equations of \citet{VesterPeter} and line data from different works: K00 for \citet{Kaspietal}; P04 for \citet{PetersonFerr}; G12 for \citet{Grier12}; B09 for \citet{Benetal}; W14 for \citet{Wangetal14}; B14 for \citet{Benetal14}; D10 for \citet{Denney}; T14 for \citet{Trev}; K07 for \citet{Kaspietal07}.} 
\label{SED10}
\end{figure*}

The BH mass and the Eddington ratio inferred with KERRBB are shown as a function of the BH spin for different values of the viewing angle $\theta_{\rm v}$; on each plot, we report different BH mass estimates from different works (listed in the caption of Fig. \ref{SED10}); a blue shaded area ($\sim 0.1$ dex) defines the confidence interval on each BH mass estimate.

\clearpage

\begin{table*}[b]
\centering
%\resizebox{\columnwidth}{!}{
\footnotesize
\begin{tabular}{lllllllllll}
\hline\\
\vspace{1mm}
Name & Redshift & Telescope & Observation Date & & &  Name & Redshift & Telescope & Observation Date\\ 
\hline   
\hline   \\
3C273 & 0.158 & FUSE & 2000 Apr 23 $^{a}$ & & & NGC5548 & 0.017 & HUT & 1995 Mar 14\\
	& & HST & 2000 Mar 16 $^{a}$ & & & & & IUE & 1995 May 16 \\
	& & KPNO & 2000 Feb 25-26 $^{a}$ & & & & & KPNO & 1985 - 1989 $^{l}$\\
\vspace{2mm}
	& & IRTF & 2007 Jan 25 $^{b}$ & & & & & \\

Ark120 & 0.033 & FUSE & 2000 Dic 31 & & & NGC7469 & 0.016 & FUSE & 2002 Dic 14\\
	& & HST & 1994 Sep - 1995 Jul & & & & & IUE & 1991 Nov 29 \\
	& & IRTF & 2007 Jan 26 $^{b}$ & & & & & KPNO & 1985 - 1989 $^{l}$\\
\vspace{2mm}
	& & & & & & & & IRTF & 2003 Oct 23 $^{f}$\\

Fairall9 & 0.047 & HUT & 1995 Mar 11 & & & PG0026+129 & 0.142 & HST & $^{c}$\\
	& & HST & 1993 Jan 21 $^{c}$ & & & & & KPNO & 1990 Oct 11 $^{e}$  \\
	\vspace{2mm}
	& & AGN Watch & 1994 - 1995 $^{d}$ & & & & & GNIRS & 2011 Aug 3 $^{m}$ \\

MRK142 & 0.045 & EUVE & 1998 May 04 & & & PG0052+251 & 0.154 & FUSE & 1999 Oct 03 $^{a}$ \\
	& & IUE & 1983 Jun 02 & & & & & HST & 1999 Oct 01 $^{a}$ \\
	& & SDSS 2007 & 2007 & & & & & KPNO & 1999 Oct $^{a}$\\
\vspace{2mm}
	& & & & & & & & GNIRS & 2011 Aug 03 $^{m}$\\

MRK290 & 0.029 & FUSE & 2007 Mar 17 & & & PG0804+761 & 0.100 & FUSE & 2002 Feb 09 \\
	& & IUE & 1985 Jan 22 & & & & &  IUE & 1986 Mar 01 \\
	& & SDSS 2007 & 2007 & & & & & KPNO & 1991 Mar $^{j}$ \\
	\vspace{2mm}	
	& & & & & & & & IRTF & 2007 Jan 24 $^{b}$\\

MRK335 & 0.026 & FUSE & 2000 Nov 21 & & & PG0844+349 & 0.064 & FUSE & 2000 Feb 20 $^{a}$ \\
	& & IUE & 1993 Set 05 & & & & & HST & 1999 Oct 21 $^{a}$ \\
	& & IRTF & 2007 Jan 25 $^{b}$ & & & & & KPNO & 2000 Feb $^{a}$\\
	\vspace{2mm}		
	& & & & & & & & IRTF & 2007 Jan 24 $^{f}$\\

MRK509 & 0.034 & FUSE & 1999 Nov 06 $^{a}$ & & & PG0953+414 & 0.234 & FUSE & 1999 Dec 30 $^{a}$ \\
	& & HST & 1992 Jun 21 $^{a}$ & & & & & HST & 2000 Feb 05 $^{a}$ \\
	& & KPNO & 1999 Dec 11 $^{a}$ & & & & & KPNO & 2000 Feb 26 $^{a}$\\
\vspace{2mm}
	& & IRTF & 2004 Jun 01 $^{f}$ & & & & & \\

MRK590 & 0.026 & IUE & 1991 Jan 15 & & & PG1211+143 & 0.081 & HUT & 1995 Mar 15 \\
	& & SDSS 2007 & 2007 & & & & & HST & $^{c}$ \\
\vspace{2mm}
	& & & & & & & & KPNO & 1991 Mar $^{j}$\\

MRK877 & 0.112 & IUE & 1993 May & & & PG1247+267 & 2.038 & HST & $^{c}$ \\
	& & KPNO & 1990 Feb 20 $^{e}$ & & & & & SDSS 2016 & \\
\vspace{2mm}
	& & & & & & & & \\

MRK1044 & 0.016 & FUSE & 2004 Jan 01 & & & PG1307+085 & 0.155 & FUSE & 1980 May 04\\
	& & IUE & 1995 Dic 21 & & & & & IUE & 2000 Jun 12 \\
	& & UKST & 2001 - 2006 $^{g}$ & & & & & KPNO & 1991 Mar $^{j}$ \\
\vspace{2mm}
	& & & 2012 - 2013 $^{h}$ & & & & & GNIRS & 2011 Aug 11 & \\

MRK1383 & 0.086 & FUSE & 2001 Mar 10 & & & PG1411+442 & 0.089 & FUSE & 2000 May 11 \\
	& & IUE & 1985 Mark 03 $^{i}$ & & & & & HST & $^{n}$ \\
	& & Steward Obs. & 1991 Mar $^{j}$ & & & & &  \\
\vspace{2mm}	
	& & & & & & & \\

MRK1501 & 0.089 & IUE & 1984 Jun 12 & & & PG1700+518 & 0.292 & HST & 1992 Aug - Dec\\
	& & KPNO & 1990 Set 18 $^{e}$ & & & & & INT & 1984 May - Jun $^{p}$\\
\vspace{2mm}
	& & & & & & & & KPNO & 1991 Mar $^{j}$\\

NGC3783 & 0.010 & FUSE & 2004 May 05 & & & PG2130+099 & 0.063 & FUSE & 2004 Nov 01 \\
	& & IUE & 1992 Jul 30 & & & & & IUE & 1985 Dic \\
	& & UKST & 2001 - 2006 $^{g}$ & & & & & KPNO & 1991 Mar $^{j}$ \\
\vspace{2mm}
	& & IRTF & 2002 Apr 25 $^{f}$\\

NGC4151 & 0.003 & FUSE & 2002 Jun 01 & & & S50836+71 & 2.172 & Palomar 200in & $^{o}$ \\
	& & IUE & 1996 Jun 09 & & & & \\
	& & Palomar 200in & 1984 Feb 15 $^{k}$ & & & & \\
\vspace{2mm}
	& & IRTF & 2002 Apr 23 $^{f}$\\

\hline
\hline   \\
\end{tabular} %}
\caption{Public Spectroscopic data for the sources of our sample (collected in the online Mikulski Archive for Space Telescopes - MAST). We report the name of the source, the redshift, the telescope, the date of observation and/or a reference work ($^{a}$ \citealt{Shang05}; $^{b}$ \citealt{Landt}; $^{c}$ \citealt{Bech}; $^{d}$ \citealt{Cast}; $^{e}$ \citealt{Boro}; $^{f}$ \citealt{Riffel}; $^{g}$ \citealt{Jones}; $^{h}$ \citealt{Wangetal14}; $^{i}$ \citealt{Kinney}; $^{j}$ \citealt{Kaspietal}; $^{k}$ \citealt{Hoetal}; $^{l}$ \citealt{Kenn}; $^{m}$ \citealt{Landt13}; $^{n}$ \citealt{Shang11}; $^{o}$ \citealt{Lawrence}; $^{p}$ \citealt{Petti}). \label{ap:alltabresres_data}}
\end{table*}

\newpage

\begin{sidewaystable*}[b]
\footnotesize
\caption{Sample of Type 1 AGNs used in this work. (1) Name of the sources ($_{a}$ from \citealt{Kaspietal}; $_{b}$ from \citealt{PetersonFerr}; $_{c}$ from \citealt{Grier12}; $_{d}$ from \citealt{Benetal}; $_{e}$ from \citealt{Wangetal14}; $_{f}$ from \citealt{Benetal14}; $_{g}$ from \citealt{Denney}; $_{h}$ from \citealt{Trev}; $_{i}$ from \citealt{Kaspietal07}); (2) (3) Logarithmic value of the spectrum peak frequency $\nu_{\rm p}$ in Hz and luminosity $\nu_{\rm p} L_{\nu_{\rm p}}$ in erg/s inferred from the SED fitting procedure. (4) (5) BH mass in solar masses and Eddington ratio $\lambda_{\rm Edd}$ computed from the SED fitting procedure; the mean uncertainties are connected to different quantities (i.e. BH spin, viewing angle, uncertainty on the spectrum peak position). (6) (7) velocity dispersion and FWHM in km/s related to the H$\beta$ line (for PG 1247+267 and S5 0836+71, these measurements are related to the CIV line). (8) Light travel time in the rest-frame expressed in days. (9) (10) Logarithmic value of the VP in solar masses computed using the velocity dispersion and the FWHM; the uncertainties are connected to $\tau_{\rm LT}$ and $\sigma_{\rm line}$(or FWHM); the systematic uncertainty is $\sim 0.5$ dex (\citealt{Peterson10}). (11) (12) Continuum luminosity at $5100 \AA$ and $1350 \AA$ in erg/s as reported in the corresponding reference paper. (13) (14) Logarithmic value of the BH mass in solar masses computed using the H$\beta$ (CIV) line FWHM and the luminosity at $5100 \AA$ ($1350 \AA$) as reported in the reference paper, through the SE relations described in \citet{VesterPeter}. SE BH mass estimates have a systematic uncertainty of $\sim 0.5$ dex (VesterOs). When a lower error is absent, it means that for that source only an upper limit is present.} 
\label{tab:res} 
%\resizebox{0.95\textwidth}{!}{

\begin{tabular}{lllllllllllllllllllllllllllllllllll}
\hline
\hline \\
Source Name & Log $\nu_{\rm p}$ & Log $\nu_{\rm p} L_{\nu_{\rm p}}$ & Log $M_{\rm BH,fit}$ & $\lambda_{\rm Edd, FIT}$ & $\sigma$ & FWHM & $\tau_{LT}$ & Log ${\rm VP}_{\sigma}$ & Log ${\rm VP}_{\rm fwhm}$ & $L_{5100\AA}$ & $L_{1350\AA}$ & Log $M_{\rm BH,H\beta}$ & Log $M_{\rm BH, CIV}$\\
 & [Hz] & [erg/s] & [$M_{\sun}$] & & [km/s] & [km/s] & [lt. days] & [$M_{\sun}$] & [$M_{\sun}$] & [erg/s] & [erg/s] & [$M_{\sun}$] & [$M_{\sun}$] \\
(1) & (2) & (3) & (4) & (5) & (6) & (7) & (8) & (9) & (10) & (11) & (12) & (13) & (14) \\
\hline 
\hline\\ 

3C273  $_{a}$ & $15.42$ & $46.39$ & $9.175 \pm 0.450$ & $0.26 \pm 0.12$ & - & $2742 \pm 58$ & $329.9^{+101.0}_{-82.9}$ & - & $8.683^{+0.134}_{-0.144}$ & $45.810$ & - & $8.691$ & -  \\

$_{b}$ & & & & & $1777 \pm 150$ & $2598 \pm 299$ & $306.8^{+68.5}_{-90.9}$ & $8.275^{+0.158}_{-0.229}$ & $8.605^{+0.182}_{-0.259}$ & $45.960$ & - & $8.719$ & - \\

\hline\\

Ark120  $_{a}$ & $15.28$ & $44.52$ & $8.520 \pm 0.450$ & $0.016 \pm 0.007$ & - & $5850 \pm 480$ & $38.6^{+5.3}_{-6.5}$ & - & $8.409^{+0.124}_{-0.154}$ & $44.140$ & - & $8.514$ & - \\

$_{b}$ & & & & & $1959 \pm 109$ & $5536 \pm 297$ & $47.1^{+8.2}_{-12.4}$ & $7.546^{+0.117}_{-0.182}$ & $8.448^{+0.115}_{-0.181}$ & $44.230$ & - & $8.511$ & - \\

 $_{b}$& & & & & $1884 \pm 48$ & $5284 \pm 203$ & $37.1^{+4.8}_{-5.4}$ & $7.408^{+0.075}_{-0.091}$ & $8.304^{+0.086}_{-0.102}$ & $44.230$ & - & $8.471$ & - \\

\hline\\

Fairall9  $_{a}$ & $15.42$ & $44.60$ & $8.280 \pm 0.450$ & $0.033 \pm 0.015$ & - & $5900 \pm 650$ & $17.1^{+3.5}_{-8.0}$ & - & $8.063^{+0.172}_{-0.375}$ & $44.140$ & - & $8.522$ & - \\

 $_{b}$& & & & & $3787 \pm 197$ & $6901 \pm 707$ & $17.4^{+3.2}_{-4.3}$ & $7.686^{+0.117}_{-0.170}$ & $8.207^{+0.158}_{-0.217}$ & $44.250$ & - & $8.713$ & - \\

\hline\\

MRK142 $_{d}$ & $15.40$ & $43.92$ & $7.980 \pm 0.450$ & $0.014 \pm 0.006$ & $859 \pm 102$ & $1368 \pm 379$ & $2.74^{+0.73}_{-0.83}$ & $5.594^{+0.200}_{-0.267}$ & $5.998^{+0.315}_{-0.438}$ & $43.540$ & - & $6.952$ & - \\

$_{e}$ & & & & & - & $1647 \pm 69$ & $6.4^{+0.8}_{-2.2}$ & - & $6.528^{+0.087}_{-0.220}$ & $43.520$ & - & $7.103$ & - \\

\hline\\

MRK290 $_{g}$ & $15.41$ & $43.95$ & $7.975 \pm 0.450$ & $0.015 \pm 0.007$ & $1606 \pm 47$ & $4270 \pm 157$ & $8.72^{+1.21}_{-1.02}$ & $6.640^{+0.081}_{-0.080}$ & $7.490^{+0.088}_{-0.087}$ & $43.110$ & - & $7.726$ & - \\

\hline\\

MRK335  $_{a}$ & $15.32$ & $44.22$ & $8.290 \pm 0.450$ & $0.014 \pm 0.006$ & - & $1260 \pm 120$ & $16.8^{+5.2}_{-3.3}$ & - & $6.715^{+0.196}_{-0.182}$ & $43.790$ & - & $7.006$ & - \\

$_{b}$ & & & & & $917 \pm 52$ & $1629 \pm 145$ & $16.8^{+4.8}_{-4.2}$ & $6.439^{+0.157}_{-0.176}$ & $6.938^{+0.183}_{-0.206}$ & $43.860$ & - & $7.264$ & - \\

$_{b}$ & & & & & $948 \pm 113$ & $1375 \pm 357$ & $12.5^{+6.6}_{-5.5}$ & $6.339^{+0.282}_{-0.362}$ & $6.662^{+0.385}_{-0.513}$ & $43.860$ & - & $7.117$ & - \\

$_{c}$ & & & & & $1293 \pm 64$ & $1025 \pm 35$ & $14.1^{+0.4}_{-0.4}$ & $6.661^{+0.054}_{-0.057}$ & $6.459^{+0.041}_{-0.043}$ & $43.700$ & - & $6.781$ & - \\

$_{e}$ & & & & & - & $1997 \pm 265$ & $10.6^{+1.7}_{-2.9}$ & - & $6.915^{+0.173}_{-0.262}$ & $43.640$ & - & $7.331$ & - \\

\hline\\

MRK509  $_{a}$ & $15.35$ & $44.68$ & $8.460 \pm 0.450$ & $0.026 \pm 0.012$ & - & $2860 \pm 120$ & $79.3^{+6.5}_{-6.2}$ & - & $8.100^{+0.070}_{-0.073}$ & $44.170$ & - & $7.908$ & - \\

$_{b}$ & & & & & $1276 \pm 28$ & $2715 \pm 101$ & $79.6^{+6.1}_{-5.4}$ & $7.401^{+0.051}_{-0.050}$ & $8.057^{+0.064}_{-0.063}$ & $44.280$ & - & $7.918$ & - \\

\hline\\

MRK590  $_{a}$ & $15.50$ & $43.18$ & $7.410 \pm 0.450$ & $0.009 \pm 0.004$ & - & $2170 \pm 120$ & $20.5^{+4.5}_{-3.0}$ & - & $7.273^{+0.133}_{-0.118}$ & $43.710$ & - & $7.438$ & - \\

$_{b}$ & & & & & $789 \pm 74$ & $1675 \pm 587$ & $20.7^{+3.5}_{-2.7}$ & $6.399^{+0.146}_{-0.146}$ & $7.052^{+0.329}_{-0.435}$ & $43.810$ & - & $7.263$ & - \\

$_{b}$ & & & & & $1935 \pm 52$ & $2566 \pm 106$ & $14.0^{+8.5}_{-8.8}$ & $7.008^{+0.229}_{-0.454}$ & $7.253^{+0.241}_{-0.467}$ & $43.810$ & - & $7.634$ & - \\

$_{b}$ & & & & & $1251 \pm 72$ & $2115 \pm 575$ & $29.2^{+4.9}_{-5.0}$ & $6.948^{+0.116}_{-0.133}$ & $7.404^{+0.276}_{-0.357}$ & $43.810$ & - & $7.466$ & - \\

$_{b}$ & & & & & $1201 \pm 130$ & $1979 \pm 386$ & $28.8^{+3.6}_{-4.2}$ & $6.907^{+0.140}_{-0.168}$ & $7.341^{+0.206}_{-0.257}$ & $43.810$ & - & $7.408$ & - \\

MRK877  $_{a}$ & $15.20$ & $44.91$ & $8.875 \pm 0.450$ & $0.017 \pm 0.008$ & - & $3880 \pm 650$ & $70.1^{+27.0}_{-36.8}$ & - & $8.312^{+0.276}_{-0.483}$ & $44.370$ & - & $8.273$ & - \\

$_{b}$ & & & & & $2626 \pm 211$ & $4718 \pm 991$ & $71.5^{+29.6}_{-33.7}$ & $7.981^{+0.218}_{-0.350}$ & $8.490^{+0.316}_{-0.482}$ & $44.480$ & - & $8.498$ & - \\

\hline 
\hline \\
\end{tabular}
\end{sidewaystable*}

\newpage

\begin{sidewaystable*}
\centering
\footnotesize
\caption{Same as Table \ref{tab:res}}\label{tab:res2}
\begin{tabular}{lllllllllllllllllllllllllllllllllll}
\hline
\hline \\
Source Name & Log $\nu_{\rm p}$ & Log $\nu_{\rm p} L_{\nu_{\rm p}}$ & Log $M_{\rm BH,fit}$ & $\lambda_{\rm Edd, FIT}$ & $\sigma$ & FWHM & $\tau_{LT}$ & Log ${\rm VP}_{\sigma}$ & Log ${\rm VP}_{\rm fwhm}$ & $L_{5100\AA}$ & $L_{1350\AA}$ & Log $M_{\rm BH,H\beta}$ & Log $M_{\rm BH, CIV}$\\
 & [Hz] & [erg/s] & [$M_{\sun}$] & & [km/s] & [km/s] & [lt. days] & [$M_{\sun}$] & [$M_{\sun}$] & [erg/s] & [erg/s] & [$M_{\sun}$] & [$M_{\sun}$] \\
(1) & (2) & (3) & (4) & (5) & (6) & (7) & (8) & (9) & (10) & (11) & (12) & (13) & (14) \\
\hline 
\hline\\ 

MRK1044 $_{e}$ & $15.50$ & $43.39$ & $7.515 \pm 0.450$ & $0.012 \pm 0.006$ & - & $1211 \pm 48$ & $4.8^{+7.4}_{-3.7}$ & - & $6.136^{+0.439}_{-0.675}$ & $43.050$ & - & $6.601$ & - \\

\hline\\

MRK1383  $_{b}$ & $15.56$ & $45.32$ & $8.360 \pm 0.450$ & $0.145 \pm 0.067$ & $3442 \pm 308$ & $6323 \pm 1295$ & $95.0^{+29.9}_{-37.1}$ & $8.340^{+0.193}_{-0.296}$ & $8.868^{+0.281}_{-0.414}$ & $44.720$ & - & $8.872$ & - \\

\hline\\

MRK1501 $_{c}$ & $15.34$ & $44.69$ & $8.485 \pm 0.450$ & $0.026 \pm 0.012$ & $3321 \pm 107$ & $5054 \pm 145$ & $15.5^{+2.2}_{-1.8}$ & $7.521^{+0.085}_{-0.082}$ & $7.886^{+0.082}_{-0.079}$ & $44.320$ & - & $8.477$ & - \\

\hline 

NGC3783  $_{a}$ & $15.40$ & $43.64$ & $7.840 \pm 0.450$ & $0.010 \pm 0.005$ & - & $4100 \pm 1160$ & $4.5^{+3.6}_{-3.1}$ & - & $7.167^{+0.472}_{-0.796}$ & $43.250$ & - & $7.761$ & - \\

$_{b}$ & & & & & $1753 \pm 141$ & $3093 \pm 529$ & $10.2^{+3.3}_{-2.3}$ & $6.785^{+0.189}_{-0.184}$ & $7.278^{+0.259}_{-0.274}$ & $43.260$ & - & $7.521$ & - \\

\hline\\

NGC4151  $_{a}$ & $15.20$ & $42.35$ & $7.595 \pm 0.450$ & $0.001 \pm 0.000$ & - & $5230 \pm 920$ & $3.0^{+1.8}_{-1.4}$ & - & $7.203^{+0.345}_{-0.441}$ & $42.860$ & - & $7.777$ & - \\

\vspace{1mm}

$_{b}$ & & & & & $1914 \pm 42$ & $4248 \pm 516$ & $3.1^{+1.3}_{-1.3}$ & $6.344^{+0.171}_{-0.255}$ & $7.036^{+0.252}_{-0.349}$ & $42.880$ & - & $7.606$ & - \\

\hline\\

NGC5548  $_{a}$ & $15.33$ & $43.52$ & $7.920 \pm 0.450$ & $0.006 \pm 0.003$ & - & $5500 \pm 400$ & $21.6^{+2.4}_{-0.7}$ & - & $8.104^{+0.107}_{-0.080}$ & $43.430$ & - & $8.106$ & - \\

$_{b}$ & & & & & $1687 \pm 56$ & $4044 \pm 199$ & $19.7^{+1.5}_{-1.5}$ & $7.037^{+0.060}_{-0.064}$ & $7.797^{+0.074}_{-0.078}$ & $43.510$ & - & $7.879$ & - \\

$_{b}$ & & & & & $2026 \pm 68$ & $5706 \pm 357$ & $16.4^{+1.2}_{-1.1}$ & $7.117^{+0.059}_{-0.060}$ & $8.016^{+0.083}_{-0.086}$ & $43.510$ & - & $8.178$ & - \\

$_{d}$ & & & & & $4270 \pm 292$ & $11177 \pm 2266$ & $4.18^{+0.86}_{-1.30}$ & $7.170^{+0.139}_{-0.223}$ & $8.006^{+0.242}_{-0.359}$ & $42.620$ & - & $8.317$ & - \\

$_{g}$ & & & & & $1822 \pm 35$ & $4849 \pm 112$ & $12.40^{+2.74}_{-3.85}$ & $6.903^{+0.103}_{-0.178}$ & $7.753^{+0.107}_{-0.182}$ & $42.660$ & - & $7.611$ & - \\

\hline\\

NGC7469  $_{a}$ & $15.40$ & $43.69$ & $7.865 \pm 0.450$ & $0.011 \pm 0.005$ & - & $3220 \pm 1580$ & $5.0^{+0.6}_{-1.1}$ & - & $7.003^{+0.396}_{-0.694}$ & $43.740$ & - & $7.796$ & - \\

$_{b}$ & & & & & $1456 \pm 207$ & $2169 \pm 459$ & $4.5^{+0.7}_{-0.8}$ & $6.268^{+0.178}_{-0.218}$ & $6.614^{+0.230}_{-0.292}$ & $43.720$ & - & $7.443$ & - \\

$_{e}$ & & & & & - & $1722 \pm 30$ & $24.3^{+4.0}_{-4.0}$ & - & $7.146^{+0.081}_{-0.093}$ & $43.560$ & - & $7.162$ & - \\

\hline\\

PG0026+129  $_{a}$ & $15.47$ & $45.31$ & $8.535 \pm 0.450$ & $0.095 \pm 0.044$ & - & $1358 \pm 91$ & $109.5^{+25.4}_{-31.6}$ & - & $7.594^{+0.147}_{-0.208}$ & $44.850$ & - & $7.601$ & - \\

$_{b}$ & & & & & $1773 \pm 285$ & $1719 \pm 495$ & $111.0^{+24.1}_{-28.3}$ & $7.831^{+0.215}_{-0.280}$ & $7.804^{+0.305}_{-0.423}$ & $45.020$ & - & $7.891$ & - \\

$_{e}$ & & & & & - & $2544 \pm 56$ & $111.0^{+24.1}_{-28.3}$ & - & $8.145^{+0.104}_{-0.147}$ & $44.840$ & - & $8.141$ & - \\

\hline

PG0052+251  $_{a}$ & $15.35$ & $45.40$ & $8.820 \pm 0.450$ & $0.060 \pm 0.028$ & - & $4550 \pm 270$ & $85.6^{+26.0}_{-26.8}$ & - & $8.537^{+0.165}_{-0.216}$ & $44.810$ & - & $8.631$ & - \\

$_{b}$ & & & & & $1783 \pm 86$ & $4165 \pm 381$ & $89.8^{+24.5}_{-24.1}$ & $7.744^{+0.146}_{-0.179}$ & $8.481^{+0.181}_{-0.219}$ & $44.960$ & - & $8.629$ & - \\

\hline\\

PG0804+761  $_{a}$ & $15.40$ & $45.34$ & $8.690 \pm 0.450$ & $0.071 \pm 0.033$ & - & $2430 \pm 42$ & $137.3^{+23.6}_{-21.8}$ & - & $8.197^{+0.084}_{-0.090}$ & $44.820$ & - & $8.091$ & - \\

$_{b}$ & & & & & $1971 \pm 105$ & $2012 \pm 845$ & $146.9^{+18.8}_{-18.9}$ & $8.045^{+0.097}_{-0.107}$ & $8.063^{+0.357}_{-0.533}$ & $44.940$ & - & $7.987$ & - \\

PG0844+349  $_{a}$ & $15.44$ & $45.00$ & $8.440 \pm 0.450$ & $0.058 \pm 0.027$ & - & $2830 \pm 120$ & $12.2^{+13.2}_{-10.3}$ & - & $7.278^{+0.355}_{-0.845}$ & $44.240$ & - & $7.934$ & - \\

$_{b}$ & & & & & $1448 \pm 79$ & $2148 \pm 612$ & $3.0^{+12.4}_{-10.0}$ & $6.087^{+0.757}_{-}$ & $6.430^{+0.928}_{-}$ & $44.350$ & - & $7.749$ & - \\

$_{e}$ & & & & & - & $2694 \pm 58$ & $32.3^{+13.7}_{-13.4}$ & - & $7.658^{+0.172}_{-0.252}$ & $44.150$ & - & $7.846$ & - \\

\hline\\

PG0953+414  $_{a}$ & $15.45$ & $45.88$ & $8.860 \pm 0.450$ & $0.167 \pm 0.077$ & - & $2723 \pm 62$ & $150.9^{+21.8}_{-26.6}$ & - & $8.337^{+0.078}_{-0.104}$ & $45.080$ & - & $8.320$ & - \\

$_{b}$ & & & & & $1306 \pm 144$ & $3002 \pm 398$ & $150.1^{+21.6}_{-22.6}$ & $7.697^{+0.149}_{-0.172}$ & $8.420^{+0.167}_{-0.194}$ & $45.220$ & - & $8.475$ & - \\

\hline\\

PG1211+143  $_{a}$ & $15.37$ & $45.20$ & $8.680 \pm 0.450$ & $0.053 \pm 0.024$ & - & $1479 \pm 66$ & $94.9^{+29.5}_{-40.5}$ & - & $7.606^{+0.155}_{-0.281}$ & $44.690$ & - & $7.595$ & - \\

$_{b}$ & & & & & $1080 \pm 102$ & $1317 \pm 138$ & $93.8^{+25.6}_{-42.1}$ & $7.328^{+0.183}_{-0.345}$ & $7.500^{+0.191}_{-0.355}$ & $44.750$ & - & $7.524$ & - \\

$_{e}$ & & & & & - & $2012 \pm 37$ & $93.8^{+25.6}_{-42.1}$ & - & $7.868^{+0.121}_{-0.275}$ & $44.660$ & - & $7.847$ & - \\

\hline 
\hline \\
\end{tabular}
\end{sidewaystable*}

\newpage

\begin{sidewaystable*}
\centering
\footnotesize
\caption{Same as Tables \ref{tab:res} - \ref{tab:res2} }\label{tab:res3}
\begin{tabular}{lllllllllllllllllllllllllllllllllll}
\hline
\hline \\
Source Name & Log $\nu_{\rm p}$ & Log $\nu_{\rm p} L_{\nu_{\rm p}}$ & Log $M_{\rm BH,fit}$ & $\lambda_{\rm Edd, FIT}$ & $\sigma$ & FWHM & $\tau_{LT}$ & Log ${\rm VP}_{\sigma}$ & Log ${\rm VP}_{\rm fwhm}$ & $L_{5100\AA}$ & $L_{1350\AA}$ & Log $M_{\rm BH,H\beta}$ & Log $M_{\rm BH, CIV}$\\
 & [Hz] & [erg/s] & [$M_{\sun}$] & & [km/s] & [km/s] & [lt. days] & [$M_{\sun}$] & [$M_{\sun}$] & [erg/s] & [erg/s] & [$M_{\sun}$] & [$M_{\sun}$] \\
(1) & (2) & (3) & (4) & (5) & (6) & (7) & (8) & (9) & (10) & (11) & (12) & (13) & (14) \\
\hline 
\hline\\ 

PG1247+267 $_{h}$ & $15.39$ & $47.58$ & $9.830 \pm 0.450$ & $0.895 \pm 0.414$ & $2104 \pm 540$ & $4568 \pm 1338$ & $142.0^{+26.0}_{-25.0}$ & $8.087^{+0.271}_{-0.342}$ & $8.760^{+0.296}_{-0.385}$ & - & $47.590$ & - & $9.882$ \\

\hline\\

PG1307+085  $_{a}$ & $15.55$ & $45.38$ & $8.410 \pm 0.450$ & $0.148 \pm 0.069$ & - & $5260 \pm 270$ & $93.5^{+39.8}_{-93.5}$ & - & $8.701^{+0.197}_{-}$ & $44.720$ & - & $8.712$ & - \\

$_{b}$ & & & & & $1820 \pm 122$ & $5058 \pm 524$ & $105.6^{+36.0}_{-46.6}$ & $7.832^{+0.184}_{-0.313}$ & $8.720^{+0.213}_{-0.348}$ & $44.880$ & - & $8.758$ & - \\

\hline\\

PG1411+442  $_{a}$ & $15.24$ & $44.90$ & $8.790 \pm 0.450$ & $0.020 \pm 0.009$ & - & $2740 \pm 110$ & $108.4^{+66.1}_{-65.2}$ & - & $8.199^{+0.241}_{-0.435}$ & $44.510$ & - & $8.040$ & - \\

$_{b}$ & & & & & $1607 \pm 169$ & $2398 \pm 353$ & $124.3^{+61.0}_{-61.7}$ & $7.795^{+0.260}_{-0.394}$ & $8.143^{+0.293}_{-0.436}$ & $44.630$ & - & $7.985$ & - \\

\hline\\

PG1700+518  $_{a}$ & $15.04$ & $45.65$ & $9.565 \pm 0.450$ & $0.019 \pm 0.009$ & - & $1970 \pm 150$ & $88.2^{+190.4}_{-88.2}$ & - & $7.823^{+0.563}_{-}$ & $45.430$ & - & $8.214$ & - \\

$_{b}$ & & & & & $1700 \pm 123$ & $1846 \pm 682$ & $251.8^{+45.9}_{-38.8}$ & $8.150^{+0.133}_{-0.138}$ & $8.222^{+0.346}_{-0.473}$ & $45.630$ & - & $8.257$ & - \\

$_{e}$ & & & & & - & $2252 \pm 85$ & $251.8^{+45.9}_{-38.8}$ & - & $8.395^{+0.105}_{-0.106}$ & $45.440$ & - & $8.335$ & - \\

\hline\\

PG2130+099  $_{a}$ & $15.39$ & $44.70$ & $8.390 \pm 0.450$ & $0.032 \pm 0.015$ & - & $3010 \pm 180$ & $177.2^{+128.2}_{-25.4}$ & - & $8.494^{+0.287}_{-0.121}$ & $44.330$ & - & $8.032$ & - \\

$_{b}$ & & & & & $1623 \pm 86$ & $2912 \pm 231$ & $158.1^{+29.8}_{-18.7}$ & $7.908^{+0.120}_{-0.102}$ & $8.416^{+0.141}_{-0.126}$ & $44.460$ & - & $8.068$ & - \\

$_{c}$ & & & & & $1825 \pm 65$ & $2097 \pm 102$ & $15.5^{+2.2}_{-1.8}$ & $7.001^{+0.088}_{-0.085}$ & $7.122^{+0.099}_{-0.097}$ & $44.150$ & - & $7.628$ & - \\

$_{e}$ & & & & & - & $2450 \pm 188$ & $31.0^{+4.0}_{-4.0}$ & - & $7.558^{+0.117}_{-0.129}$ & $44.130$ & - & $7.753$ & - \\

\hline\\

S50836+71 $_{i}$ & $15.46$ & $47.22$ & $9.510 \pm 0.450$ & $0.816 \pm 0.377$ & - & $9700 \pm 0$ & $188.0^{+27.0}_{-37.0}$ & - & $9.536^{+0.058}_{-0.095}$ & - & $47.050$ & - & $10.250$ \\

\hline 
\hline \\
\end{tabular} 
\end{sidewaystable*}

%--------------------------------------------------------
%--------------------------------------------------------

\newpage

\begin{figure*}[b]
\centering
\hskip -0.2 cm
\includegraphics[width=0.505\textwidth]{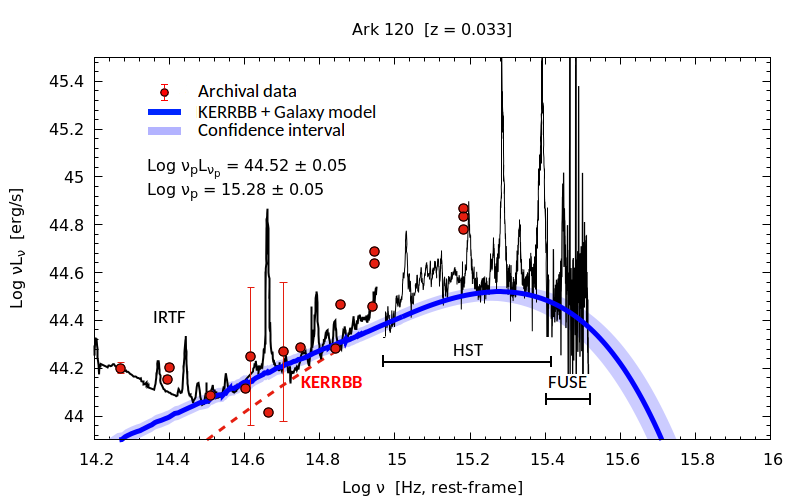}\includegraphics[width=0.505\textwidth]{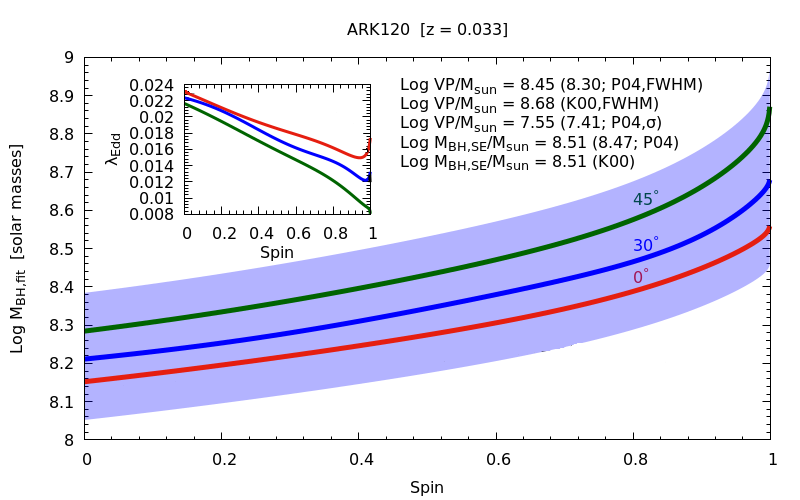}
\caption{Other BH mass estimates are from \citet{Haas} (Log ${\rm VP}(\sigma_{\rm line}) / M_{\odot} = 7.53^{+0.13}_{-0.19}$) and \citet{Du} (Log $\rm{VP}(\rm{FWHM})/M_{\odot} = 8.12^{+0.08}_{-0.09}$). The galaxy emission is necessary to obtain a satisfactory fit at Log $\nu /{\rm Hz} = 14.4 - 14.8$. The spectrum rise at Log $\nu /{\rm Hz} < 14.5$ is caused by the IR emission of the dusty torus.} 
\label{SED11}
\end{figure*}

\begin{figure*}[b]
\centering
\hskip -0.2 cm
\includegraphics[width=0.505\textwidth]{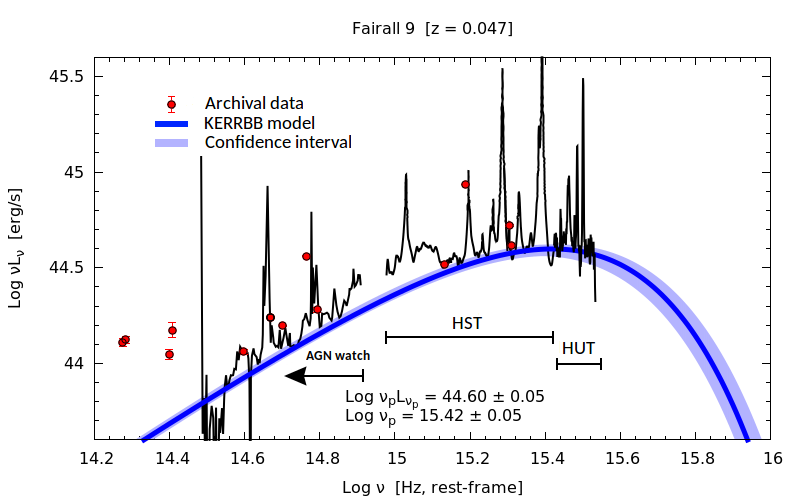}\includegraphics[width=0.505\textwidth]{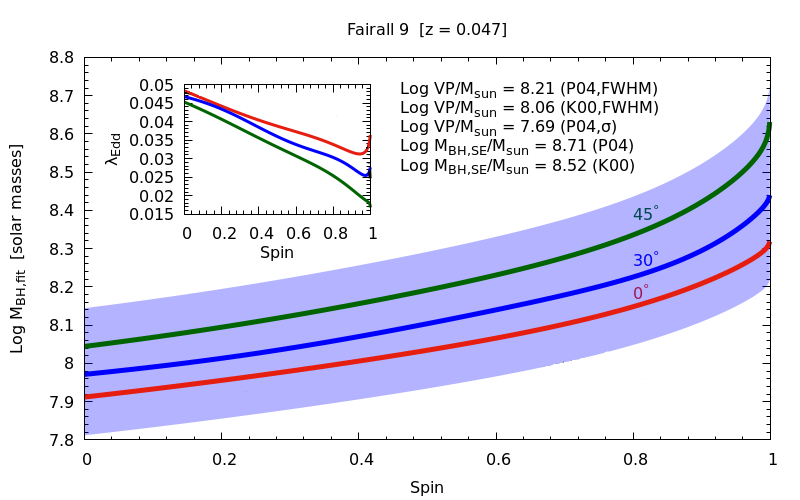}
\caption{Fit of the source Fairall 9.} 
\label{SED12}
\end{figure*}

\newpage 

\begin{figure*}[b]
\centering
\hskip -0.2 cm
\includegraphics[width=0.505\textwidth]{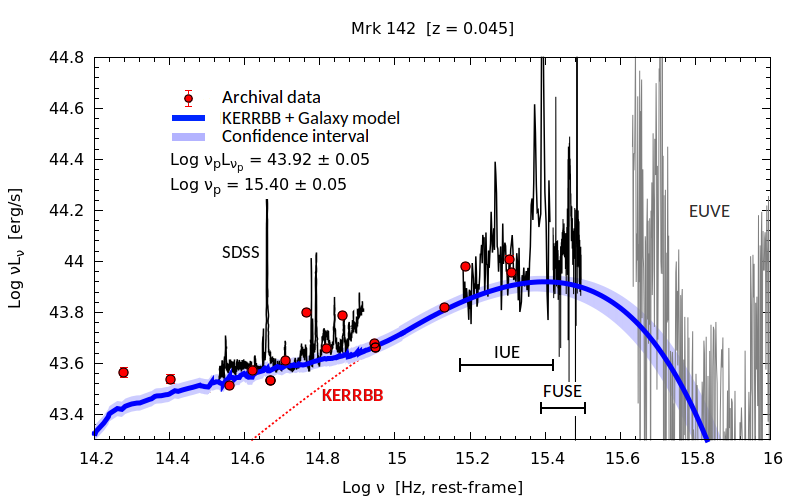}\includegraphics[width=0.505\textwidth]{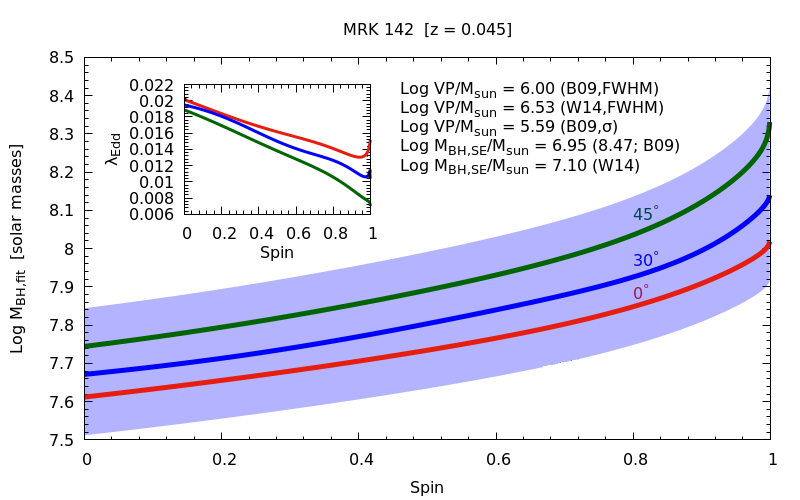}
\caption{Spectroscopic data in the EUV region are very noisy and represented in gray to give an idea of the emission at large frequencies. Another BH mass estimate is from \citet{Li18} (Log ${\rm VP}(\sigma_{\rm line}) / M_{\odot} = 6.23^{+0.26}_{-0.45}$).} 
\label{SED13}
\end{figure*}

\begin{figure*}[b]
\centering
\hskip -0.2 cm
\includegraphics[width=0.505\textwidth]{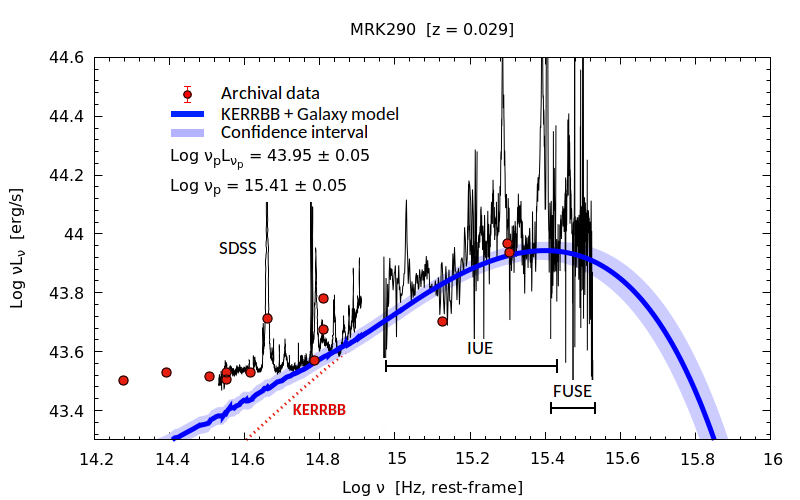}\includegraphics[width=0.505\textwidth]{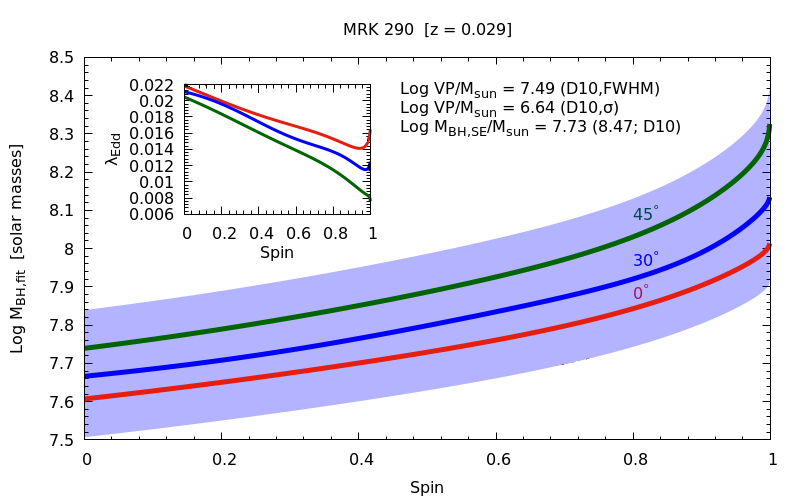}
\caption{Spectrum rise at Log $\nu /{\rm Hz} < 14.5$ caused by the IR emission of the dusty torus.} 
\label{SED14}
\end{figure*}

\begin{figure*}[b]
\centering
\hskip -0.2 cm
\includegraphics[width=0.505\textwidth]{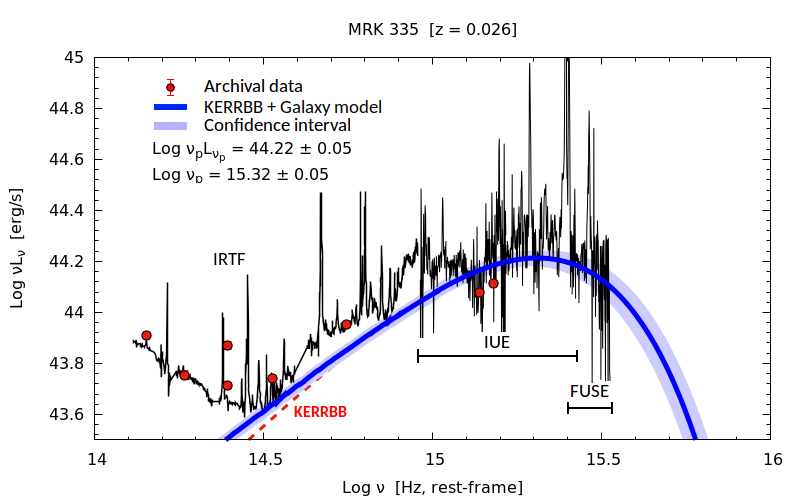}\includegraphics[width=0.505\textwidth]{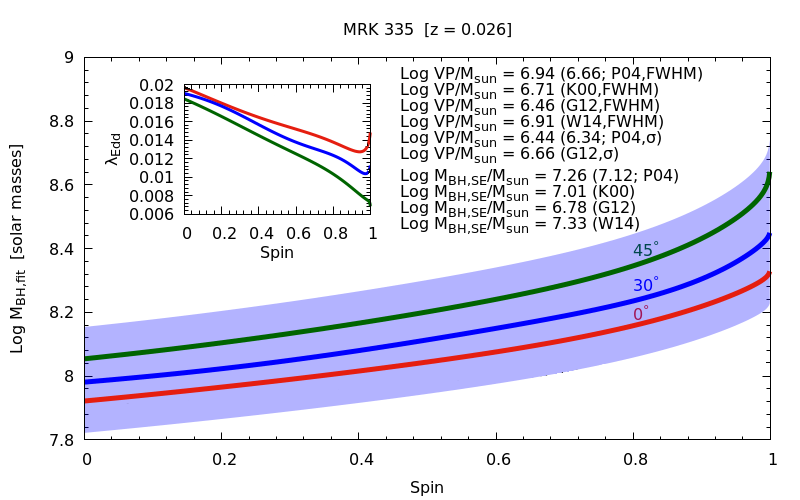}
\caption{HST data around the Ly$\alpha$-CIV region are also available, in good agreement with IUE data. The spectrum rise at Log $\nu /{\rm Hz} < 14.4$ is caused by the IR emission of the dusty torus. Other BH mass estimates are from \citet{Haas} (Log $M / M_{\odot} = 6.45^{+0.14}_{-0.22}$) and \citet{Grier18} (Log $M / M_{\odot} = 7.25 \pm 0.10$).} 
\label{SED15}
\end{figure*}

\newpage

\begin{figure*}[b]
\centering
\hskip -0.2 cm
\includegraphics[width=0.505\textwidth]{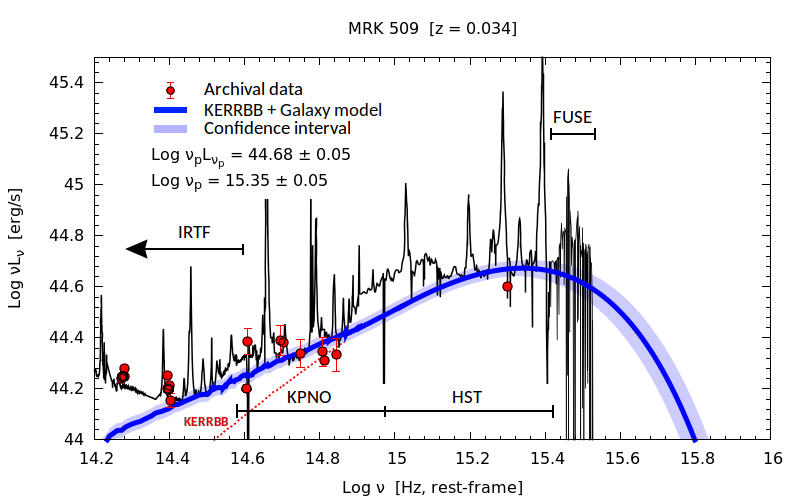}\includegraphics[width=0.505\textwidth]{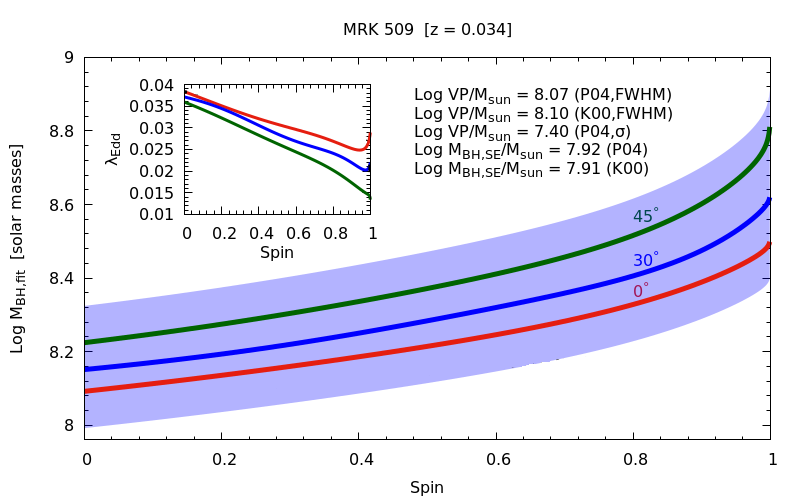}
\caption{Spectrum rise at Log $\nu /{\rm Hz} < 14.5$ caused by the IR emission of the dusty torus.} 
\label{SED16}
\end{figure*}

\begin{figure*}[b]
\centering
\hskip -0.2 cm
\includegraphics[width=0.505\textwidth]{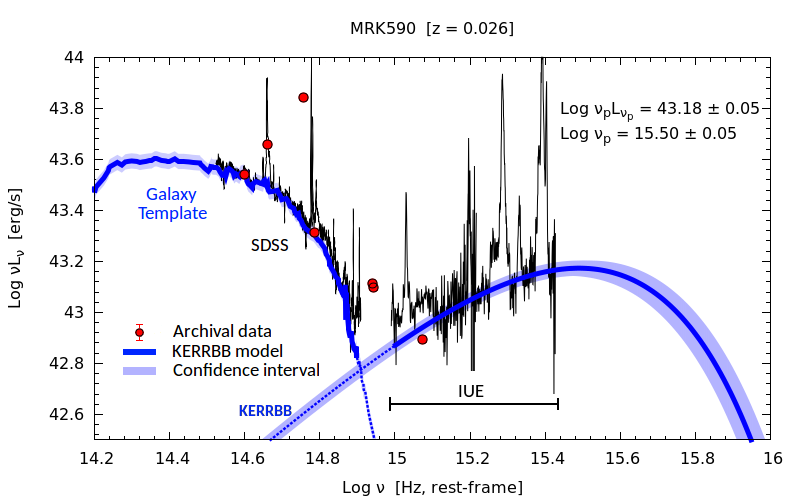}\includegraphics[width=0.505\textwidth]{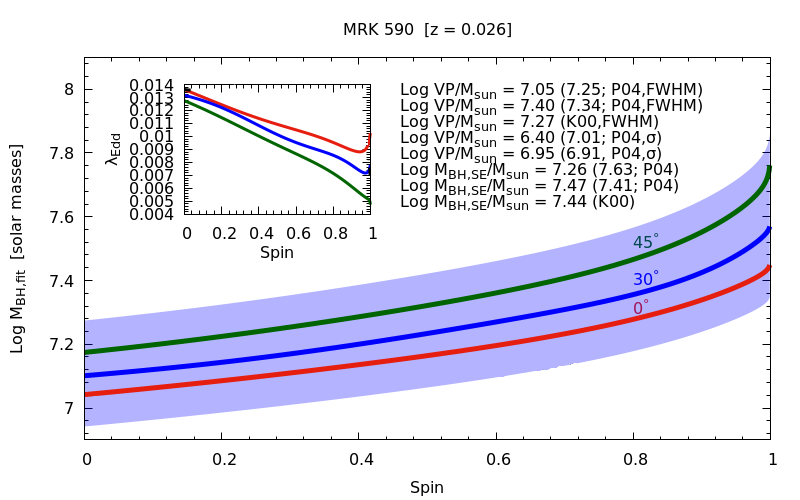}
\caption{Prominent host-galaxy emission required in the fitting procedure; for a satisfactory fit, the AD emission has to be cut at around Log $\nu /{\rm Hz} \sim 15$ (i.e., the AD size is smaller than $10^6 R_{\rm g}$ as implemented in KERRBB). In the range Log $\nu /{\rm Hz} = 14.9 - 15$, the Balmer continuum can describe the rise of spectroscopic data (not represented for clarity). For this source, the peak is not visible but we used the curvature at smaller frequencies to obtain an estimate of the position.} 
\label{SED17}
\end{figure*}

\begin{figure*}[b]
\centering
\hskip -0.2 cm
\includegraphics[width=0.505\textwidth]{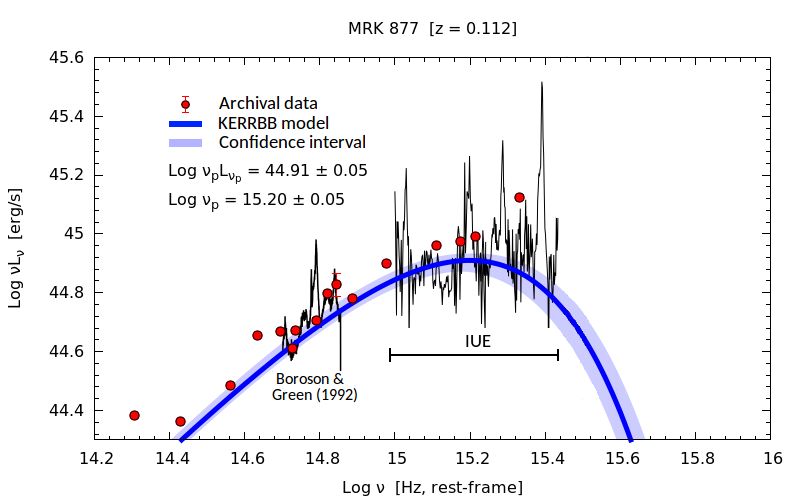}\includegraphics[width=0.505\textwidth]{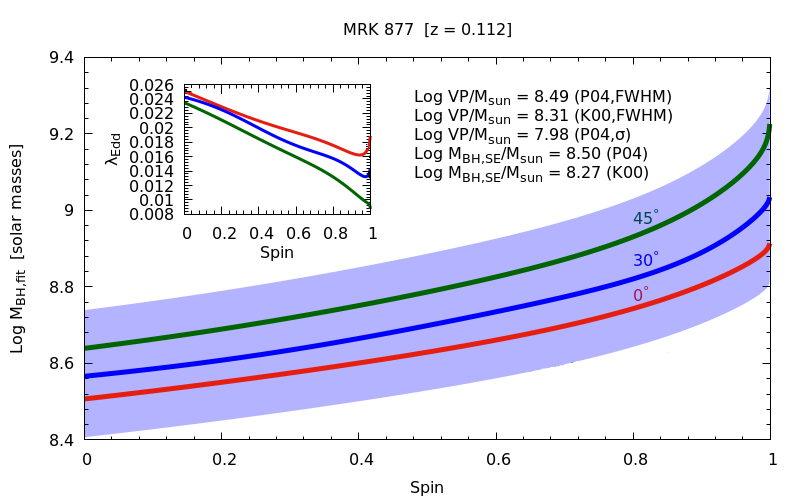}
\caption{Some absorption features present at Log $\nu /{\rm Hz} \sim 15.1 - 15.2$; nevertheless, they do not interfere in the fitting procedure.} 
\label{SED18}
\end{figure*}

\newpage

\begin{figure*}[b]
\centering
\hskip -0.2 cm
\includegraphics[width=0.505\textwidth]{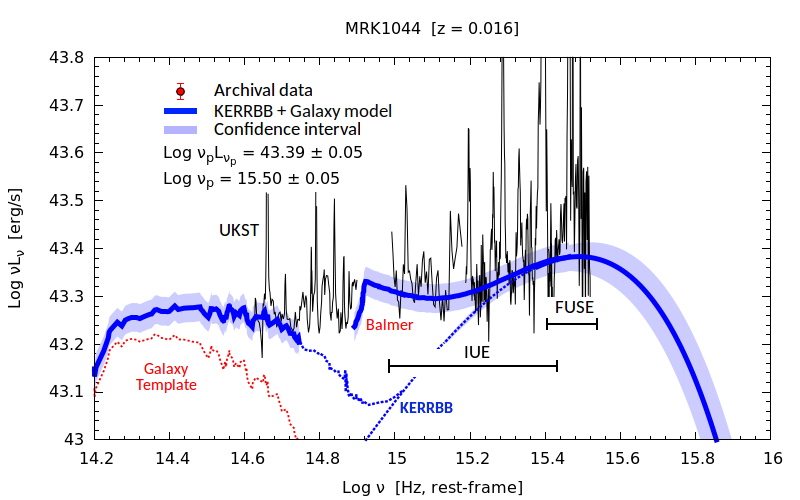}\includegraphics[width=0.505\textwidth]{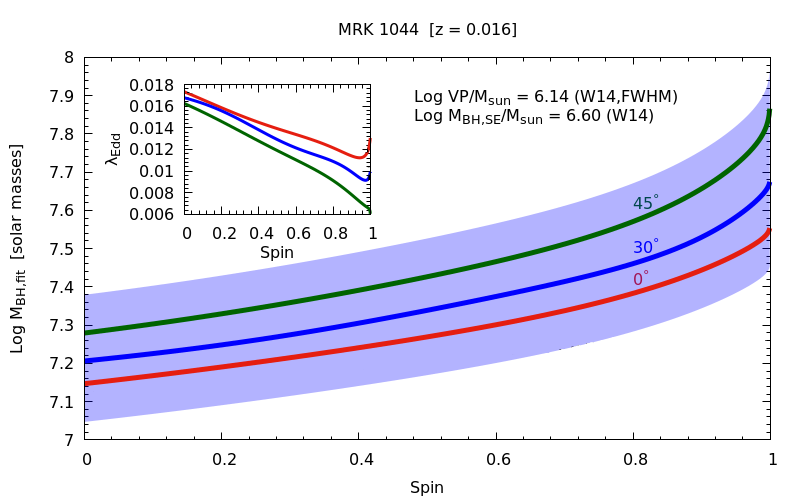}
\caption{The quality of spectroscopic data is low but nonetheless the fit was rather satisfactory. The Balmer continuum is shown in order to visualize the rise in the spectrum at Log $\nu /{\rm Hz} \sim 15$.} 
\label{SED19}
\end{figure*}

\begin{figure*}[b]
\centering
\hskip -0.2 cm
\includegraphics[width=0.505\textwidth]{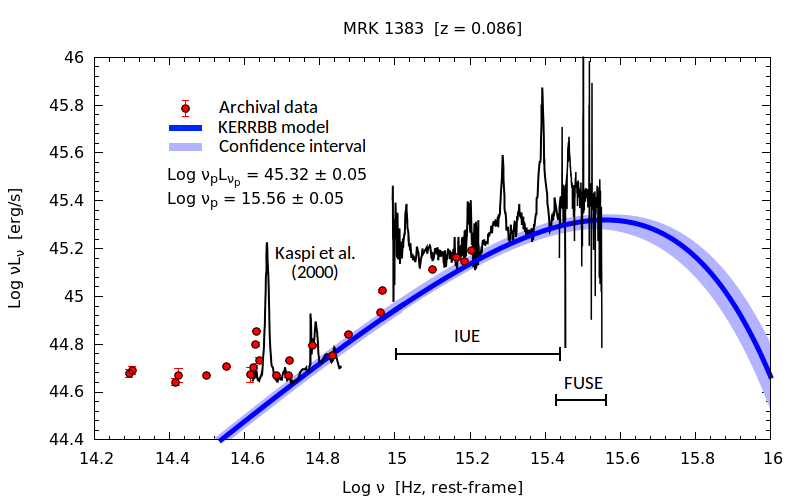}\includegraphics[width=0.505\textwidth]{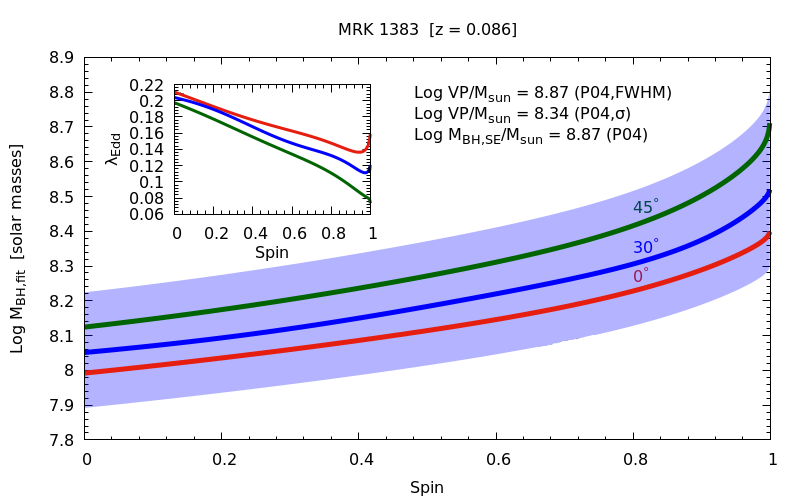}
\caption{Fit of the source MRK 1383.} 
\label{SED20}
\end{figure*}

\begin{figure*}[b]
\centering
\hskip -0.2 cm
\includegraphics[width=0.505\textwidth]{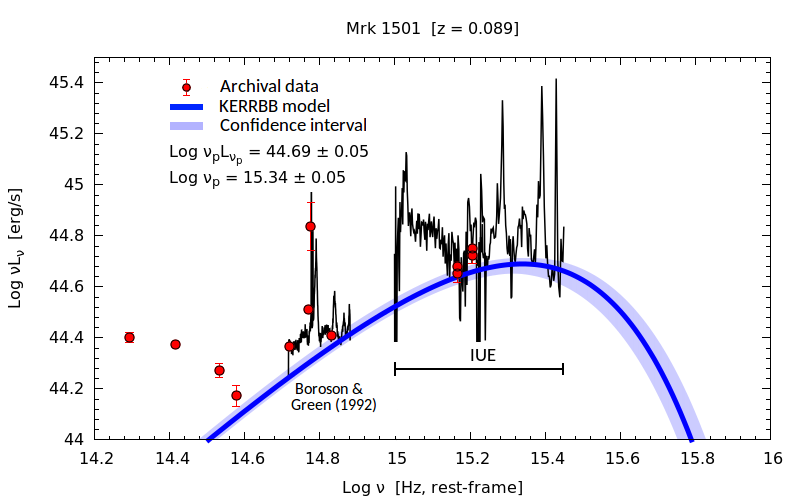}\includegraphics[width=0.505\textwidth]{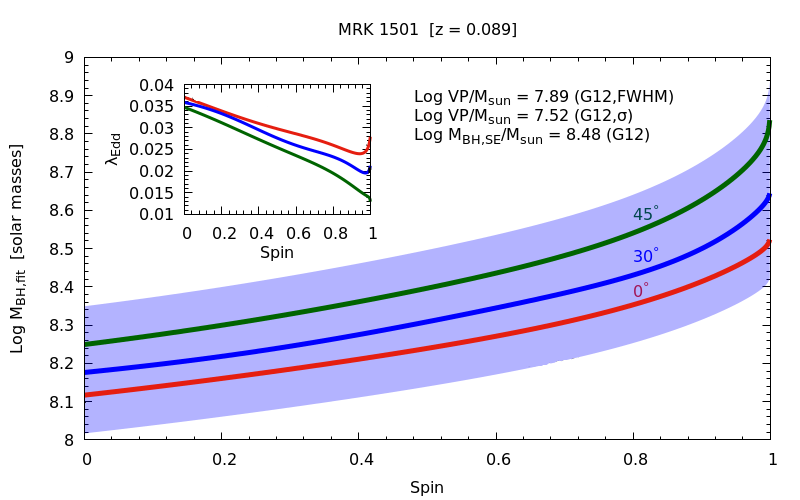}
\caption{Balmer continuum added to obtain a better visualization of the best fit. Spectroscopic data are not excellent (also without FUV) but are good enough to localize the spectrum peak. The other BH mass estimate shown is from \citet{Grier18} (Log $M / M_{\odot} = 7.84^{+0.14}_{-0.19}$).} 
\label{SED21}
\end{figure*}

\newpage

\begin{figure*}[b]
\centering
\hskip -0.2 cm
\includegraphics[width=0.505\textwidth]{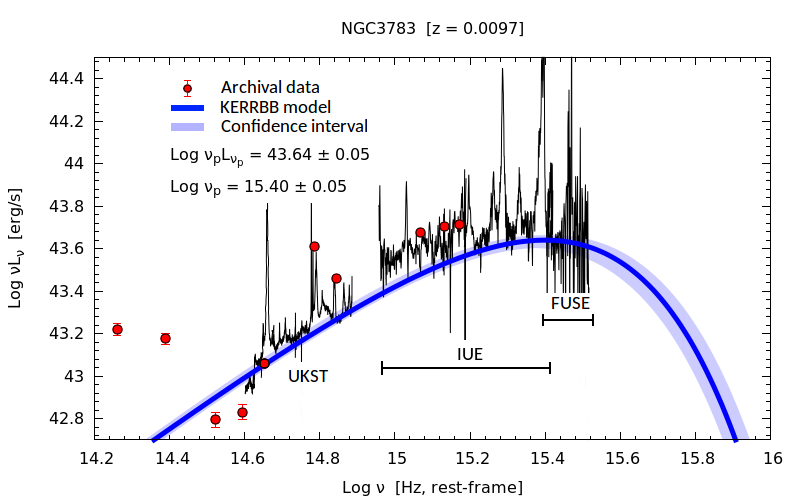}\includegraphics[width=0.505\textwidth]{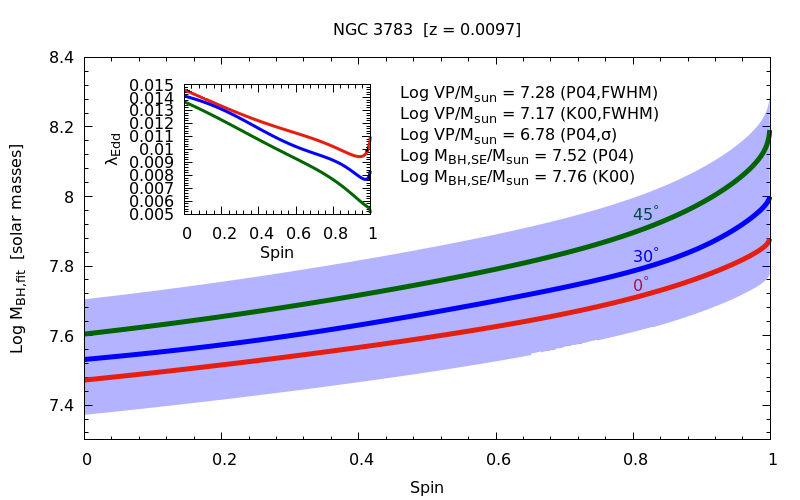}
\caption{Another BH mass estimate from \citet{Kolla} (Log $M / M_{\odot} = 7.47$).} 
\label{SED23}
\end{figure*}

\begin{figure*}[b]
\centering
\hskip -0.2 cm
\includegraphics[width=0.505\textwidth]{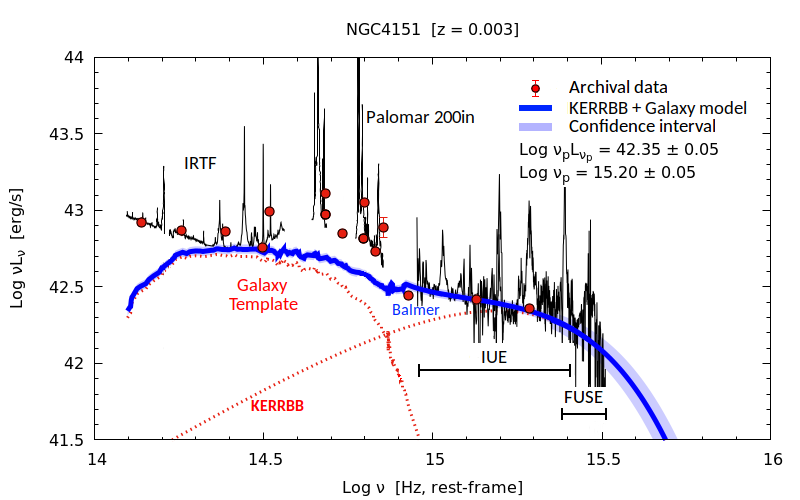}\includegraphics[width=0.505\textwidth]{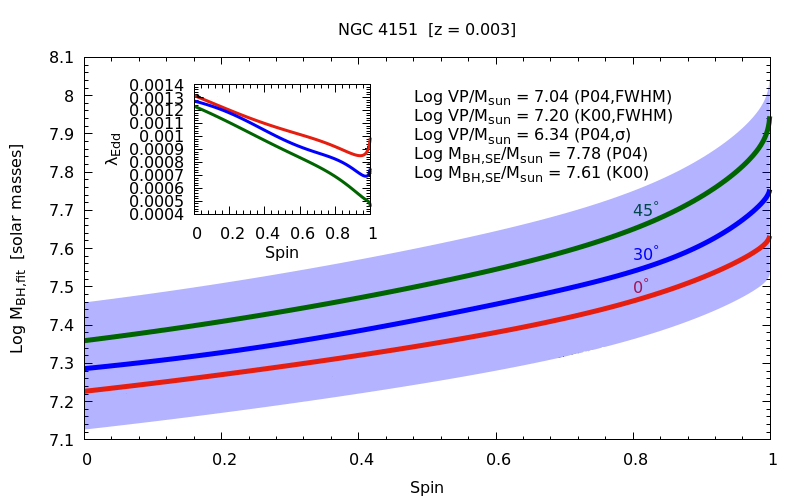}
\caption{Balmer continuum added to obtain a better visualization of the best fit. Spectrum rise at Log $\nu /{\rm Hz} < 14.3$ caused by the IR emission of the dusty torus.} 
\label{SED24}
\end{figure*}

\begin{figure*}[b]
\centering
\hskip -0.2 cm
\includegraphics[width=0.505\textwidth]{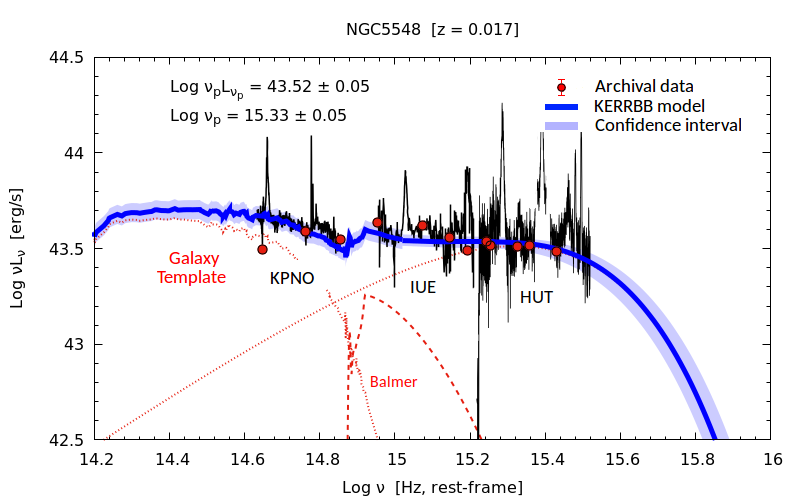}\includegraphics[width=0.505\textwidth]{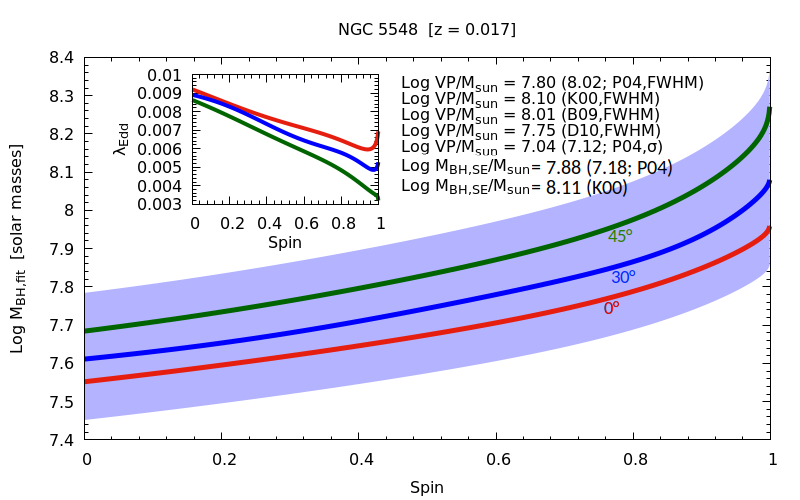}
\caption{Balmer continuum added to obtain a better visualization of the best fit. IUE, HST, and HUT data were smoothed in order to have a clearer spectrum. Other BH mass estimates are from \citet{Derosa18} (Log ${\rm VP}(\sigma_{\rm line}) / M_{\odot} = 6.74 \pm 0.06$) and \citet{Kolla} (Log $M / M_{\odot} = 7.83$).} 
\label{SED26}
\end{figure*}

\newpage

\begin{figure*}[b]
\centering
\hskip -0.2 cm
\includegraphics[width=0.505\textwidth]{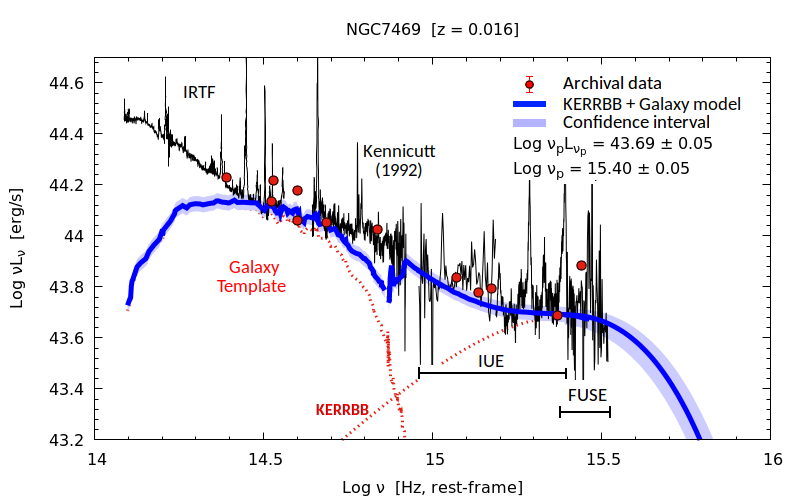}\includegraphics[width=0.505\textwidth]{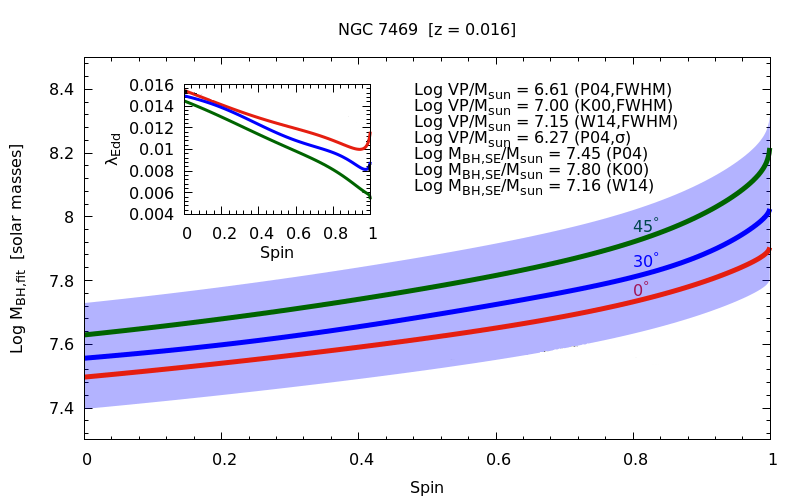}
\caption{Balmer continuum added to obtain a better visualization of the best fit. Spectrum rise at Log $\nu /{\rm Hz} < 14.5$ caused by the IR emission of the dusty torus. Another BH mass estimate is from \citet{Kolla} (Log $M / M_{\odot} = 7.09$).} 
\label{SED27}
\end{figure*}

\begin{figure*}[b]
\centering
\hskip -0.2 cm
\includegraphics[width=0.505\textwidth]{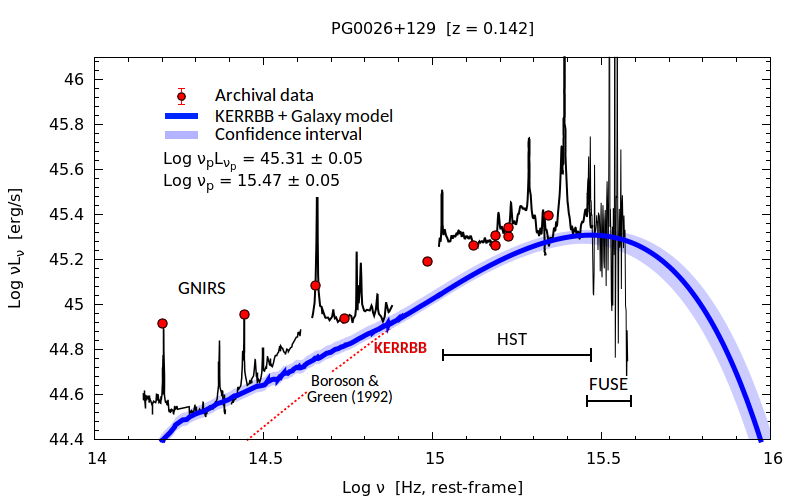}\includegraphics[width=0.505\textwidth]{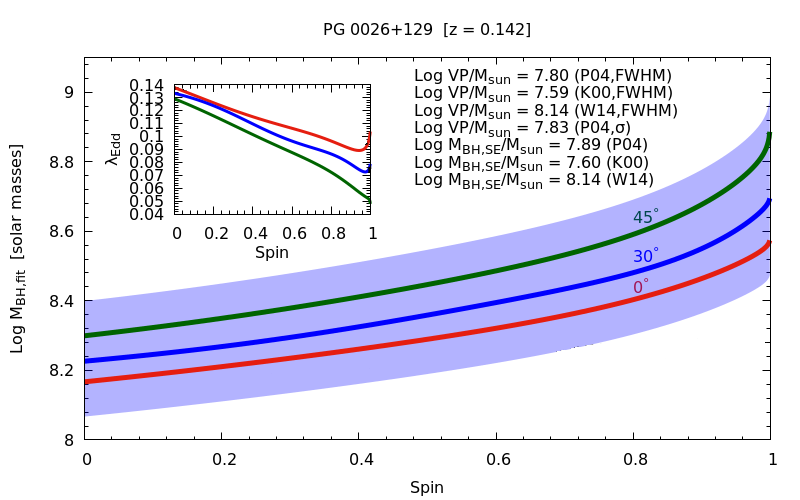}
\caption{Fit of the source PG0026+129.} 
\label{SED28}
\end{figure*}

\begin{figure*}[b]
\centering
\hskip -0.2 cm
\includegraphics[width=0.505\textwidth]{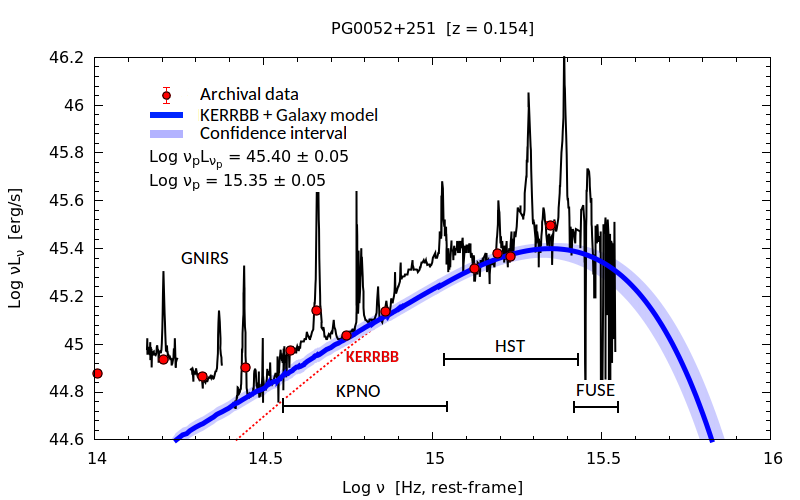}\includegraphics[width=0.505\textwidth]{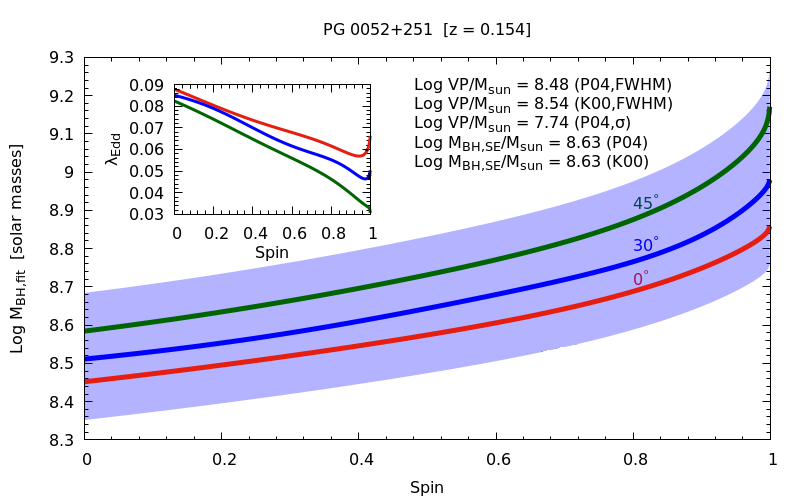}
\caption{Spectrum rise at Log $\nu /{\rm Hz} < 14.5$ caused by the IR emission of the dusty torus.} 
\label{SED29}
\end{figure*}

\newpage

\begin{figure*}[b]
\centering
\hskip -0.2 cm
\includegraphics[width=0.505\textwidth]{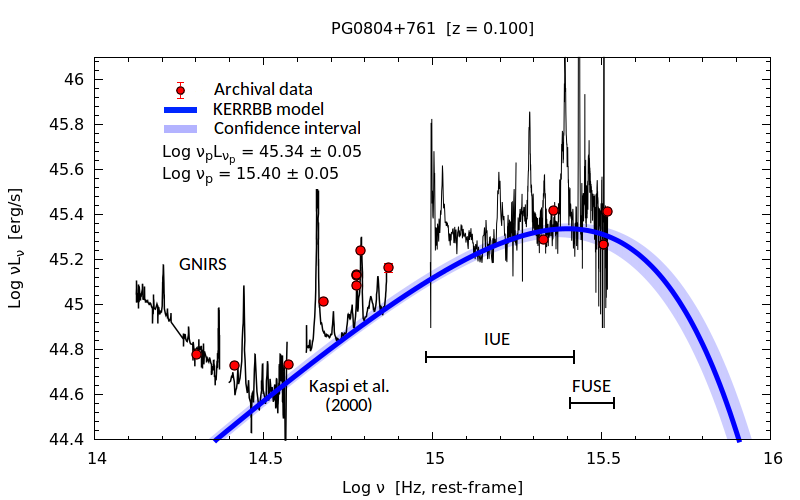}\includegraphics[width=0.505\textwidth]{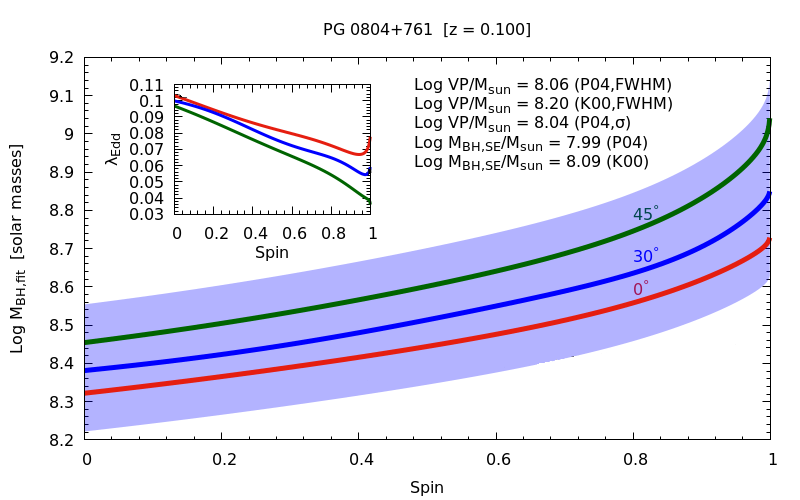}
\caption{Spectrum rise at Log $\nu /{\rm Hz} < 14.5$ caused by the IR emission of the dusty torus.} 
\label{SED30}
\end{figure*}

\begin{figure*}[b]
\centering
\hskip -0.2 cm
\includegraphics[width=0.505\textwidth]{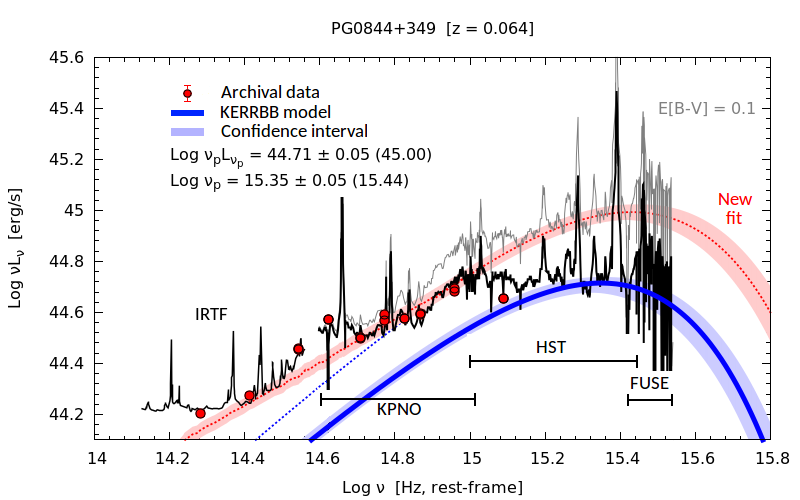}\includegraphics[width=0.505\textwidth]{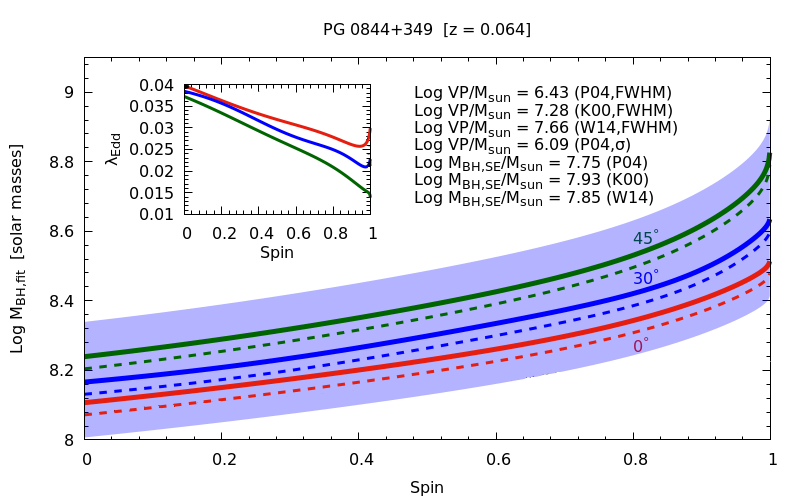}
\caption{For this source, adding a galaxy template to the blue line fit or a prominent Balmer continuum does not lead to a satisfactory fit. Instead, assuming an intrinsic reddening of the source (corrected by assuming $E[B-V] = 0.1$ mag), the new fit (red line) describes the AGN continuum for Log $\nu > 14.8$ (for smaller frequencies, we added the host-galaxy emission). Right panel: Thick lines corresponds to the first fit (blue line on the left panel); dashed lines to the second one (red line on the left panel).} 
\label{SED31}
\end{figure*}

\begin{figure*}[b]
\centering
\hskip -0.2 cm
\includegraphics[width=0.505\textwidth]{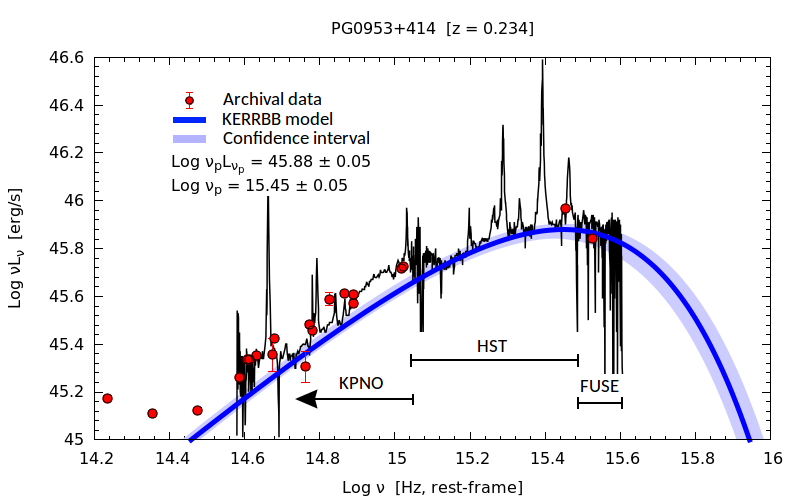}\includegraphics[width=0.505\textwidth]{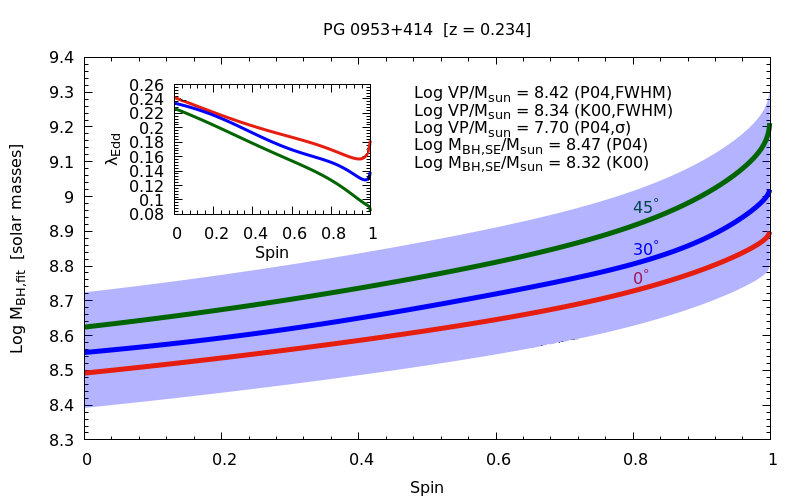}
\caption{Fit of the source PG0953+414.} 
\label{SED32}
\end{figure*}

\newpage

\begin{figure*}[b]
\centering
\hskip -0.2 cm
\includegraphics[width=0.505\textwidth]{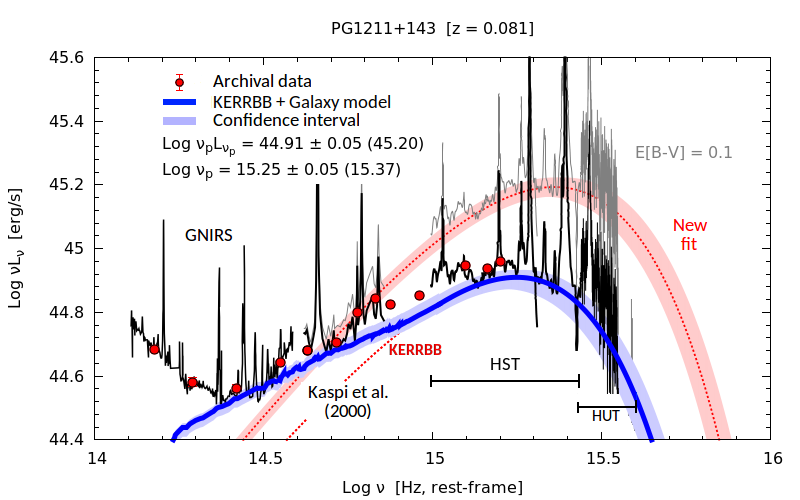}\includegraphics[width=0.505\textwidth]{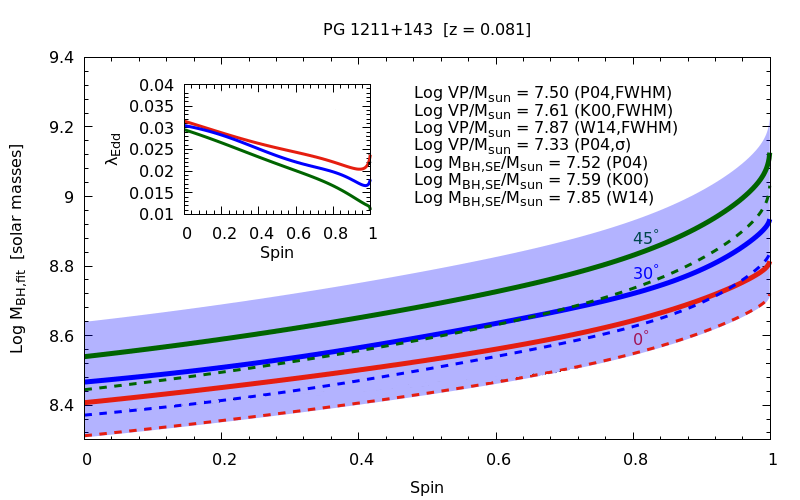}
\caption{For this source, the first satisfactory fit (blue line) is given by KERRBB+host galaxy. The spectrum rise at Log $\nu /{\rm Hz} < 14.5$ is caused by the IR emission of the dusty torus. Assuming an intrinsic reddening of the source (corrected by assuming $E[B-V] = 0.1$ mag), the new fit (red line) describea the AGN continuum for Log $\nu > 14.8$, with no need for the host-galaxy emission. Right panel: Thick and dashed lines correspond to the first and second fits, respectively (blue and red lines on the left panel, respectively). } 
\label{SED33}
\end{figure*}

\begin{figure*}[b]
\centering
\hskip -0.2 cm
\includegraphics[width=0.505\textwidth]{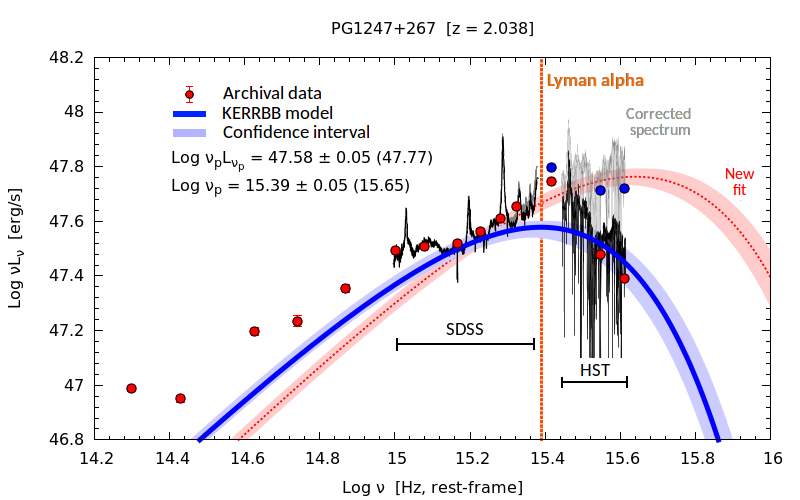}\includegraphics[width=0.505\textwidth]{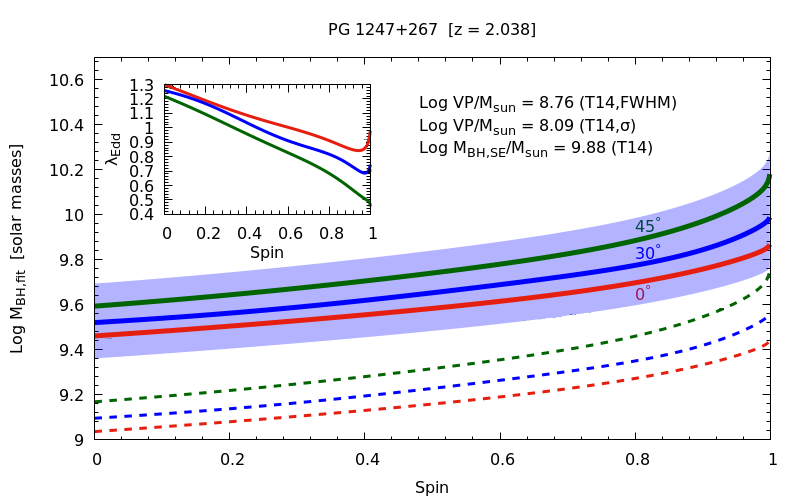}
\caption{For this source, we tried to correct the spectrum emission from IGM absorption at large frequencies by following \citet{Madau} and \citet{HaaMad}. Right panel: Thick and dashed lines correspond to the first and second fits, respectively (blue and red lines on the left panel, respectively).} 
\label{SED34}
\end{figure*}

\begin{figure*}[b]
\centering
\hskip -0.2 cm
\includegraphics[width=0.505\textwidth]{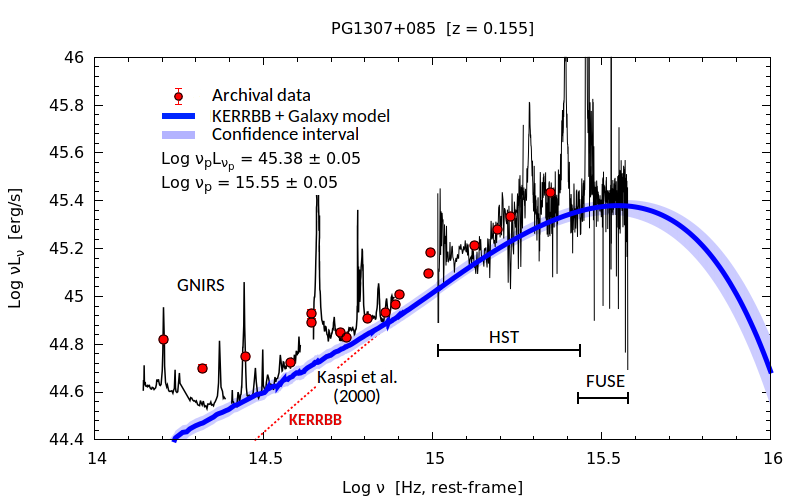}\includegraphics[width=0.505\textwidth]{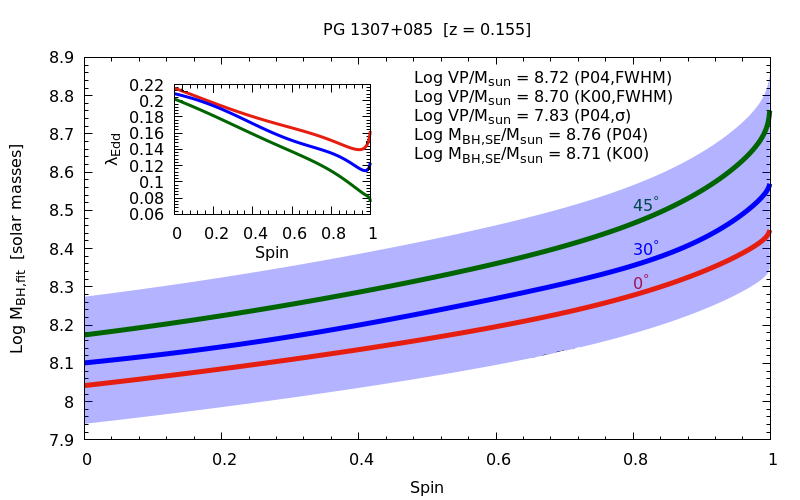}
\caption{Fit of the source PG1307+085.} 
\label{SED35}
\end{figure*}

\newpage

\begin{figure*}[b]
\centering
\hskip -0.2 cm
\includegraphics[width=0.505\textwidth]{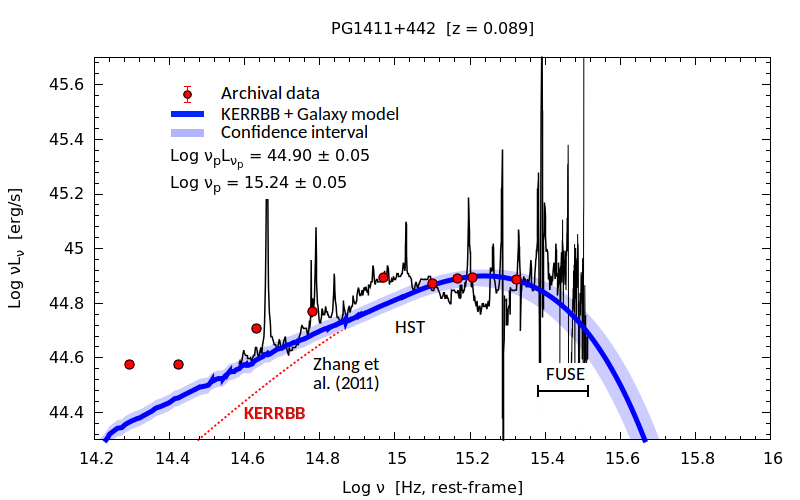}\includegraphics[width=0.505\textwidth]{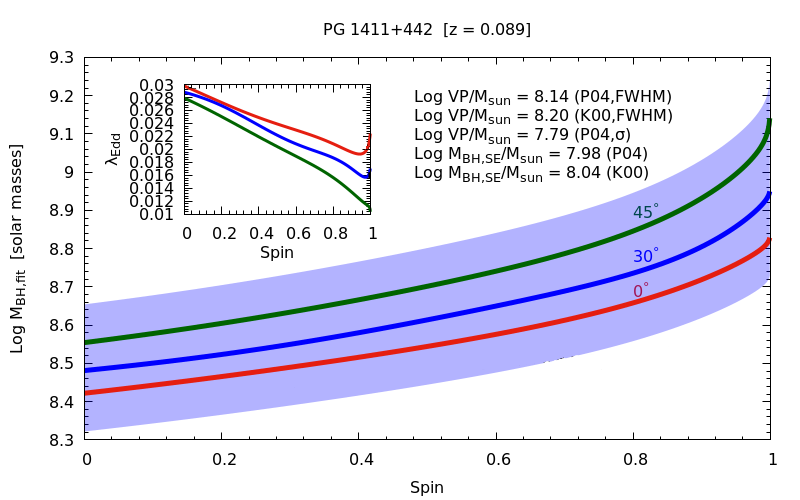}
\caption{Some intrinsic absorption is present in the data but does not affect the fit.} 
\label{SED36}
\end{figure*}

\begin{figure*}[b]
\centering
\hskip -0.2 cm
\includegraphics[width=0.505\textwidth]{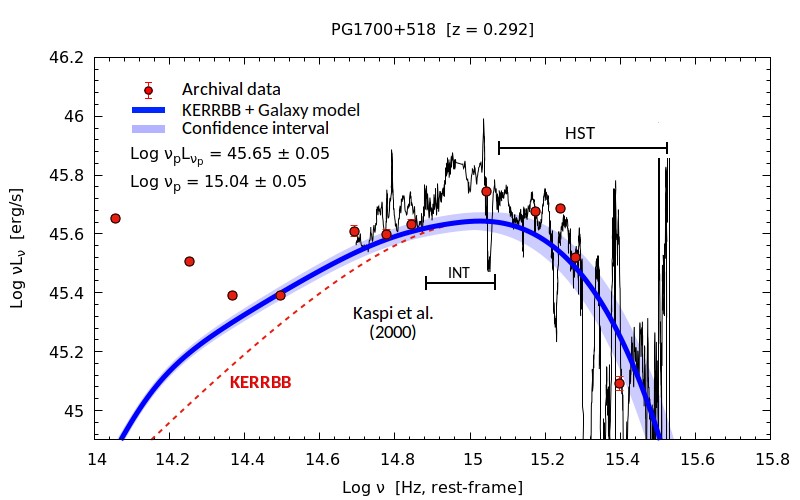}\includegraphics[width=0.505\textwidth]{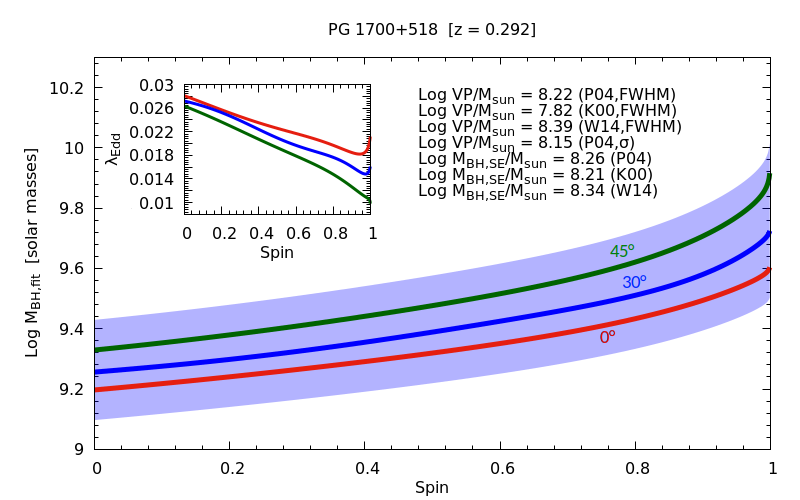}
\caption{The quality of spectroscopic data at large frequencies is low; we predicted a mass larger by a factor of $\sim 10$ with respect to RM and SE. Dust absorption could be the cause of the decreasing flux at Log $\nu / {\rm Hz} > 15$. We added the galaxy emission to the fit in order to obtain a better description of the optical continuum.} 
\label{SED37}
\end{figure*}

\begin{figure*}[b]
\centering
\hskip -0.2 cm
\includegraphics[width=0.505\textwidth]{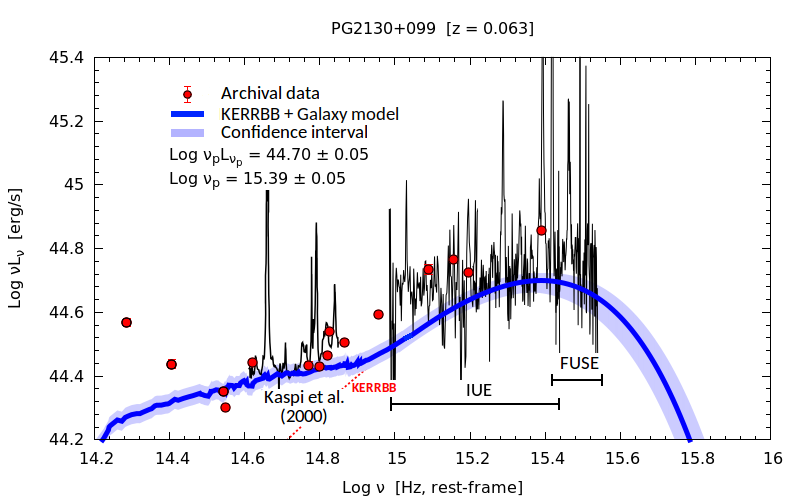}\includegraphics[width=0.505\textwidth]{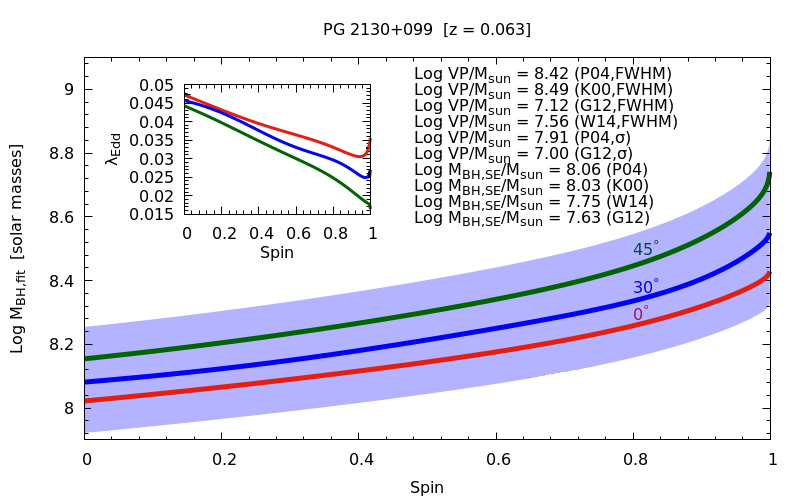}
\caption{Another BH mass estimate from \citet{Grier18} (Log $M / M_{\odot} = 6.92^{+0.24}_{-0.23}$).} 
\label{SED38}
\end{figure*}

\newpage

\begin{figure*}[b]
\centering
\hskip -0.2 cm
\includegraphics[width=0.505\textwidth]{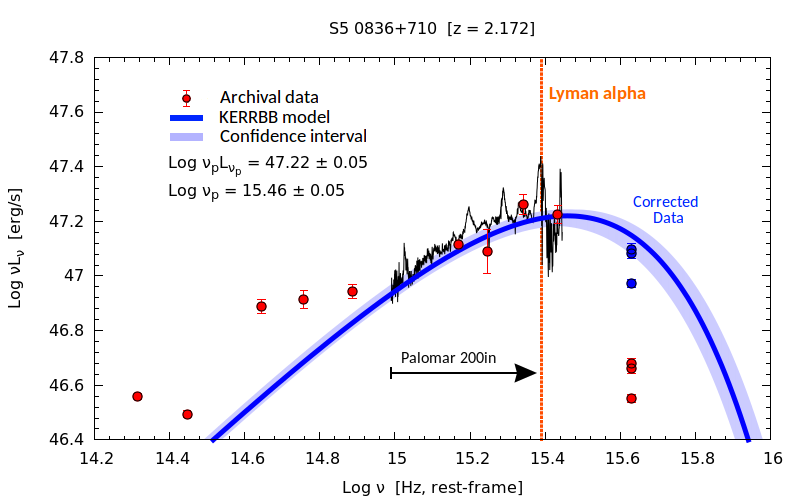}\includegraphics[width=0.505\textwidth]{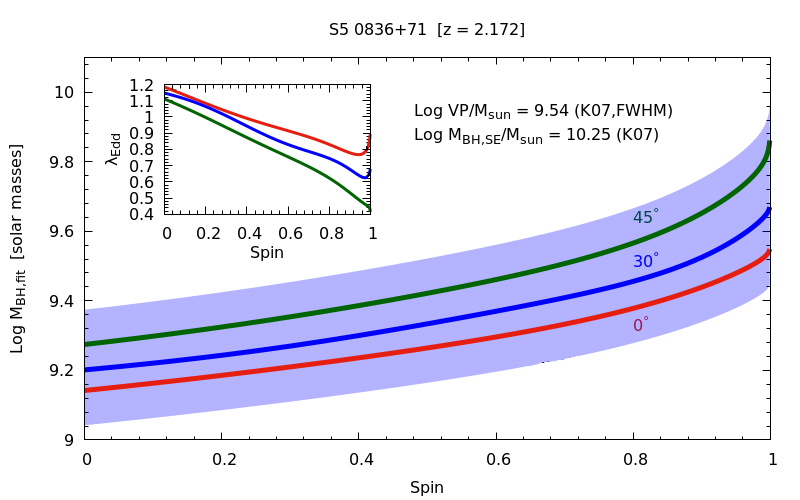}
\caption{Also for this source, we tried to correct photometric data from IGM absorption at large frequencies by following \citet{Madau} and \citet{HaaMad}; new data (blue dots) are consistent with the best fit.} 
\label{SED39}
\end{figure*}

%--------------------------------------------------
%--------------------------------------------------
%--------------------------------------------------
%--------------------------------------------------

%\newpage

%\twocolumn

\section{KERRBB equations}\label{AP_B}

The relativistic model KERRBB (\citealt{Lietal}) describes the emission produced by a thin disk around a Kerr BH. The authors included all relativistic effects such as frame-dragging, gravitational redshift, Doppler beaming and light bending. C18 build an analytic approximation of the KERRBB disk emission features considering a hardening factor equal to $1$, no limb-darkening effect, and including the self-irradiation: in the case of a face-on disk, C18 found analytic expressions to compute the BH mass $M$ and accretion rate $\dot{M}$ by fitting a given SED for different spin values. The spectrum peak $\nu_{\rm p}$ and luminosity $\nu_{\rm p} L_{\nu_{\rm p}}$ are
\begin{equation}\label{nupeak}
	\frac{\nu_{\rm p}}{\text{[Hz]}} = \mathcal{A} \left[ \frac{\dot{M}}{M_{\odot, \text{yr}^{-1}}} \right]^{1/4} 
\left[\frac{M}{10^9 M_{\odot}} \right]^{-1/2} g_{\rm 1}(a, \theta_{\rm v}),
\end{equation}
\begin{equation}\label{nulnupeak}
	\frac{\nu_{\rm p} L_{\nu_{\rm p}}}{\text{[erg/s]}} = \mathcal{B} \left[ \frac{\dot{M}}{M_{\odot, \text{yr}^{-1}}} \right] \cos \theta_v\ g_{\rm 2}(a, \theta_{\rm v}),
\end{equation}
	
\noindent where Log $\mathcal{A} = 15.25$, Log $\mathcal{B} = 45.66$, and $g_{1, 2}$ is a function depending on the viewing angle of the system $\theta_{\rm v}$ with respect to our line of sight and the BH spin $a$, containing all the relativistic modifications. The observed disk luminosity is $L^{\rm obs}_{\rm d} = \mathcal{F}(\theta_{\rm v}, a)\ \eta(a) \dot{M} c^2 \sim 2 \nu_{\rm p} L_{\nu_{\rm p}}$, where $\eta$ is the disk spin-dependent radiative efficiency and $\mathcal{F}$ depends on the BH spin and the viewing angle.\footnote{The total disk luminosity $L_{\rm d} = \eta \dot{M} c^2$ differs from the observed disk luminosity (i.e. the frequency integrated AD luminosity, $L^{\rm obs}_{\rm d} = \int L_{\nu} d \nu$) because of relativistic effect described by $\mathcal{F}$ (for details see C18).} $\mathcal{F}$, and $g_{1,2}$ have an analytical form reported in C18 and C19. 

From Eqs. \ref{nupeak} - \ref{nulnupeak}, the BH mass and the Eddington ratio (defined as $\lambda_{\rm Edd} = \eta \dot{M} c^2 / \mathcal{C} (M/M_{\odot})$, where $\mathcal{C} = 1.26 \cdot 10^{38}$ erg/s) are:
\begin{equation}\label{mass}
	\frac{M}{M_{\odot}} = \mathcal{D}\ \frac{[g_{\rm 1}(a, \theta_{\rm v})]^2}{\sqrt{\cos \theta_{\rm v}\ g_{\rm 2}(a, \theta_{\rm v})}}\ \frac{\sqrt{\nu_{\rm p} L_{\nu_{\rm p}}}}{\nu^2_{\rm {\rm p}}},
\end{equation}
\begin{equation}\label{Eddratio}
	\lambda_{\rm Edd} = \mathcal{E}\ \frac{\eta(a)}{g^2_{\rm 1}(a, \theta_{\rm v}) \sqrt{\cos \theta_{\rm v}\ g_{\rm 2}(a, \theta_{\rm v})}} \nu^2_{\rm p} \sqrt{\nu_{\rm p} L_{\nu_{\rm p}}},
\end{equation}

\noindent where Log $\mathcal{D}=16.67$ and Log $\mathcal{E}=-53.675$.

\section{Comparison with other AD models}\label{AP:AGNSED}

\subsection{Shakura and Sunyaev}

The most simple AD model is the one described by Shakura and Sunyaev (\citealt{SS}) which refers to an optically thick and geometrically thin disk around a SMBH without relativistic effects. The overall spectrum shape is similar to the KERRBB one and, as shown in \citet{Caldero}, C18 and C19, assuming the same peak position, the Shakura and Sunyaev model corresponds to a particular KERRBB solution with a precise spin value (e.g. Fig. 3 in C19). 

\subsection{SLIMBH}

As noted before, for very luminous QSOs the thin disk approximation breaks down due to the non-negligible disk vertical structure, and KERRBB results are no longer trustworthy. Another relativistic thin AD model is SLIMBH (\citealt{Abretal}; \citealt{Sad09}; \citealt{SadwAbra09}; \citealt{SadwAbra}) which accounts for the vertical structure of the disk and more appropriate for bright disks (e.g., see \citealt{Kora}). Given the similarity in shape of the two models, C19 showed that, for a fixed spectrum peak position, the differences between the two models (in terms of BH masses and Eddington ratios) are less than a factor of $\sim 1.2$ (see Figs. 3 - 4 in C19).

\subsection{AGNSED}\label{APP:AGNSED}

\begin{figure*}
\centering
\hskip -0.2 cm
\includegraphics[width=0.49\textwidth]{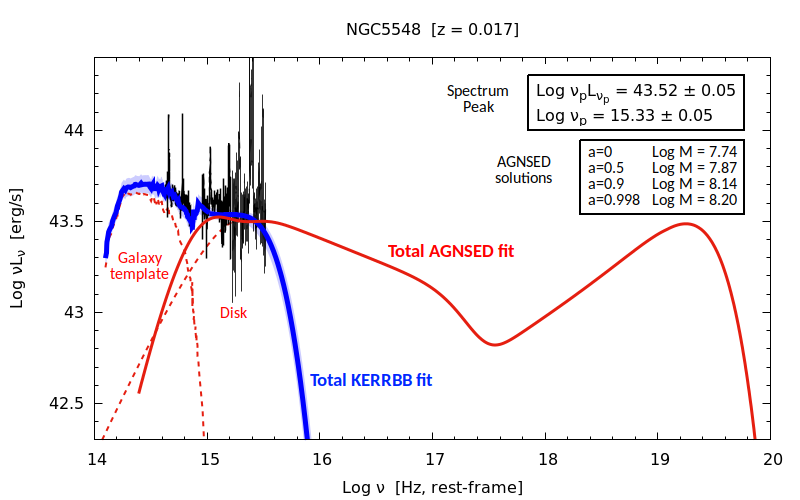}\includegraphics[width=0.49\textwidth]{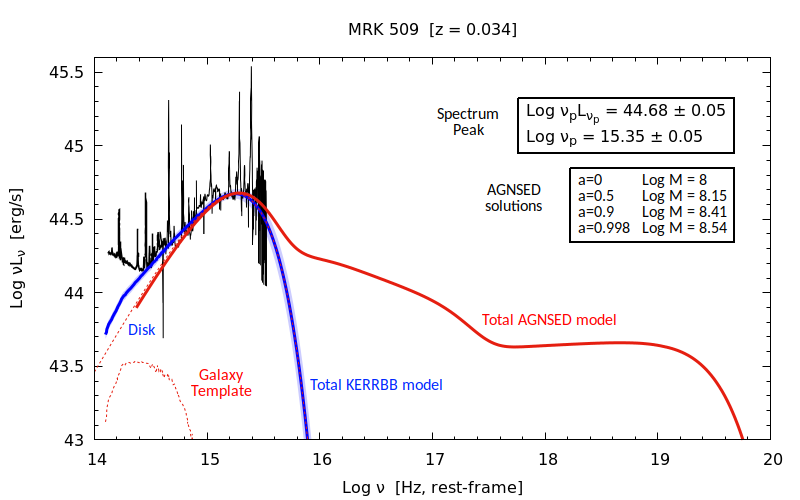} \caption{SED modeling of the sources NGC5548 (left panel) and MRK509 (right panel). The red line is the modeling of KD18 (disk + Corona) and the blue line is the modeling adopted in this work (disk + Galaxy). As KERRBB, the AGNSED modeling is degenerate and by changing some of its parameters (i.e., BH mass, spin, Eddington ratio, X-ray corona size) appropriately, it is possible to reproduce the same SED (keeping the corona slopes constant as reported in KD18): on the plot, we report some AGNSED BH mass solutions (in solar masses) for different spin values, along with the spectrum peak position (frequency and luminosity, in Hz and erg/s respectively).}
\label{AGNSED}
\end{figure*}

\begin{figure*}
\centering
\hskip -0.2 cm
\includegraphics[width=0.465\textwidth]{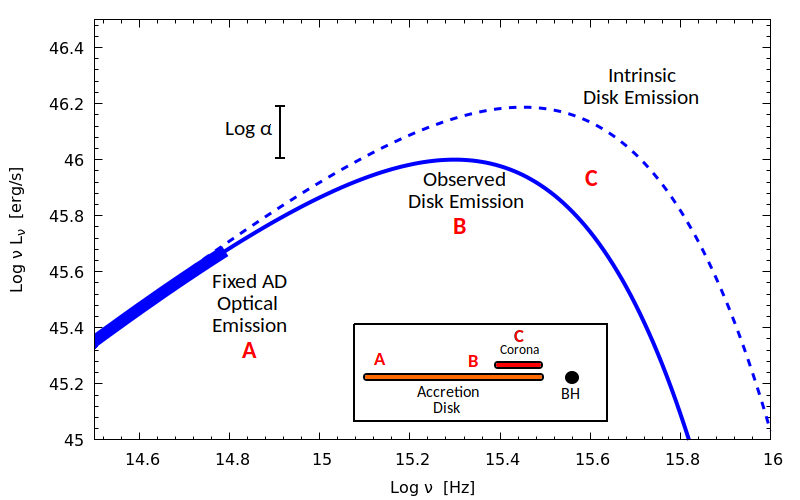} \includegraphics[width=0.50\textwidth]{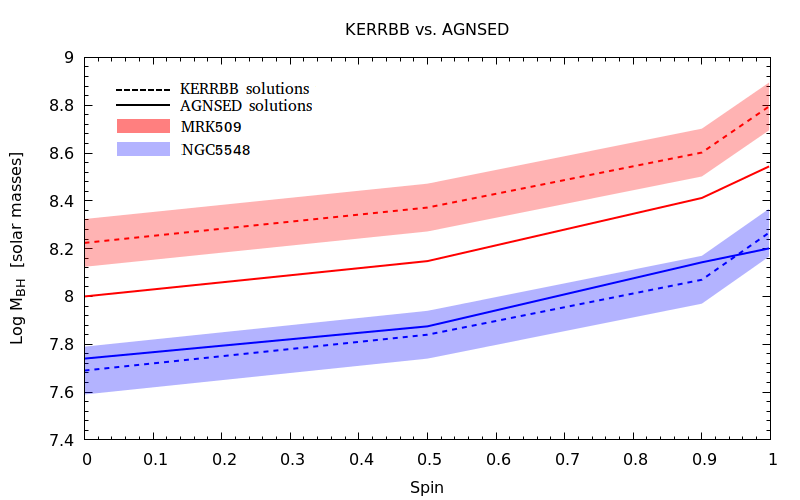}\caption{Left panel: Scheme used to explain the difference between KERRBB and AGNSED results. The observed AD luminosity is dimmer than the intrinsic one by a factor $\alpha$ because a compact corona (located in the innermost part of the disk) scatters part of the disk radiation (C). The optical part of the disk (A, i.e. the emission produced by the most distant annuli of the AD) is fixed because the corona does not cover its radiation. The intrinsic disk emission can be approximately reconstructed by taking into account the factor $\alpha$ and by fixing the low-frequency emission; the peak of the emission (B) is produced by the AD parts closer to the corona. Right panel: Comparison between AGNSED BH mass solutions (thick lines) as a function of the spin, and KERRBB ones (dashed lines) for NGC5548 (blue) and MRK509 (red). All these solutions describe the same SED plotted in Fig. \ref{AGNSED} for $\theta_{\rm v} = 45^{\circ}$. For NGC5548, the results of both models are consistent while for MRK509, masses differ by a factor $\lsim 0.2$ dex. The shaded area represents the uncertainty of $\sim 0.1$ dex (on average) linked to the spectrum peak position (see Sect. \ref{5.1}). }
\label{AGNSED22}
\end{figure*}

\citet{Kora} discussed several issues related to the black body-like AD models for AGNs: one of those is related to the soft X-ray excess observed in many objects and described by modeling an X-ray corona above the disk which scatters part of its radiation. 

The AGNSED model (KD18) is a relativistic model that describes the SED of AGNs by also taking into account the contribution of a X-ray corona located above the disk. The authors followed \citet{NovTho} to describe the AD emission: relativistic effects as the ones implemented in KERRBB (e.g., gravitational redshift, light bending, self-irradiation) are not included even though they may have a significant effect on both the disk and the X-ray corona emissions; however, for low spin values, those effects should have a minor weight on the results, especially for small viewing angles. 

In order to compare the BH masses inferred with such a model and those estimated with KERRBB, we used the sources NGC5548 and MRK509, both present in KD18 and our sample.

In KD18, the authors fitted the UV-X SED of those sources assuming a viewing angle $\theta_{\rm v} = 45^{\circ}$ (the other parameters of the model are listed in their Tab. 2, assuming an outer disc emission). As in KERRBB, AGNSED is degenerate and by changing some of its parameters (i.e., BH mass, spin, Eddington ratio, corona size) appropriately, it is possible to reproduce the same SED (see Figs. \ref{AGNSED}-\ref{AGNSED22}). 

Figure \ref{AGNSED22} (right panel) shows the comparison between the BH mass estimates from KERRBB and AGNSED as a function of the BH spin: for NGC5548, results from both models are compatible while for MRK509, KERRBB BH masses are larger than the ones found with AGNSED by a factor $\lsim 0.2$ dex. We argue that for NGC5548 the absence of relativistic effects in AGNSED is balanced out by modeling a large corona above the AD, leading to the same BH mass. Instead, for MRK509, the smaller X-ray corona leads to different results because it does not compensate the differences between the two models.

If KERRBB disk luminosity is corrected from the corona coverage, then the same value of the BH mass can be found with both models. We tested this possibility by considering this simple approach: we assumed that the observed disk luminosity $L^{\rm obs}_{\rm d}$ (and so $\nu_{\rm p} L_{\nu_{\rm p}}$) is dimmer than the intrinsic one $L^{\rm obs}_{\rm d, \alpha}$ by a factor $\alpha <1$: this means that $(1-\alpha)$ of the disk radiation is scattered by the corona (i.e., the inner disk does not contribute to the observed emission; see Fig. \ref{AGNSED22} left panel); assuming that the corona has a compact structure, the optical part of the spectrum produced by the outer part of the disk must not change due to the scattering and must keep the same luminosity. 

We find the corrected spectrum luminosity $\nu_{\rm p, \alpha} L_{\nu, \alpha}$ and frequency $\nu_{\rm p, \alpha}$ by taking into account the correction factor $\alpha$ and by keeping the same luminosity at lower frequencies:
		\begin{equation}\label{alpha}
			\nu_{\rm p, \alpha} L_{\nu, \alpha} \sim \frac{\nu_{\rm p} L_{\nu_{\rm p}}}{\alpha} \qquad \qquad \nu_{\rm p, \alpha} \sim \nu_{\rm p}\ \alpha^{-0.75},
		\end{equation}
		
\noindent where the exponent is derived assuming that the luminosity at Log $\nu \lsim 14.8$ is constant.\footnote{Following the work of \citealt{DavLao}, from their Eq. 5, we have $\nu L_{\nu, \rm opt} \propto (\dot{M} M)^{2/3}$, where $\nu L_{\nu, \rm opt}$ is the luminosity at Log $\nu /{\rm Hz} \sim 14.8$. By using Eqs. \ref{nupeak} - \ref{mass}, it is possible to write $\nu L_{\nu, \rm opt} \propto (\nu_{\rm p} L_{\nu_{\rm p}}) \nu^{-4/3}_{\rm p}$: this luminosity must be constant before and after the correction with the factor $\alpha$ therefore we can write $(\nu_{\rm p} L_{\nu_{\rm p}}) \nu^{-4/3}_{\rm p} \sim (\nu_{\rm p, \alpha} L_{\nu,\alpha}) \nu^{-4/3}_{\rm p,\alpha}$. From this latter, knowing that $\nu_{\rm p, \alpha} L_{\nu, \alpha} = \nu_{\rm p} L_{\nu_{\rm p}} / \alpha$, we finally have $\nu_{\rm p, \alpha} \sim \nu_{\rm p} \alpha^{-3/4}$.} 
		
The factor $\alpha$ mimics the actual correction of the disk inner emission when this latter is cut at a certain distance from the SMBH: using the relativistic AD model described by \citet{NovTho}, as an example, we considered the case with a non-spinning BH where the disk inner boundary is set to $20 R_{\rm g}$ (where $R_{\rm g} = GM / c^2$ is the gravitational radius), similar to the corona size adopted by KD18 for MRK509; we found that the disk peak frequency and luminosity are reduced by a factor of $\sim 0.14$ and $\sim 0.17$ dex, respectively; such corrections can be found for $\alpha \sim 0.7$. From Eqs. \ref{alpha}, the corrected mass is smaller by factor $\sim \alpha$.

Assuming that the corona scatters $\lsim 30 \%$ of the disk radiation (e.g., \citealt{Sazo}; \citealt{LuRi}), the BH mass is reduced by a factor $\lsim 0.15$ dex. We find that, for MRK509, a correction with $\alpha \sim 0.7$ is needed in order to make KERRBB results compatible with those found with AGNSED, while for NGC5548, no correction is necessary because the results of both models are already compatible (see Figs. \ref{AGNSED22}).

This simple analysis showed that both models are partially consistent even though the physical background is different. Nonetheless, despite the rather good results obtained with the $\alpha$ correction, we warn the reader about these findings: additional uncertainties (e.g. corona geometry and size) could play an important role in estimating BH masses. Moreover, even without any correction, compatibility between the results from KERRBB, RM (SE) is still rather good (Fig. \ref{MASS_SEVP_comp_fit_ecc}).

%----------------------------------------------------------------------------------------------------------------
%----------------------------------------------------------------------------------------------------------------

\end{document}